\documentclass[aps,pra,twocolumn,,superscriptaddress,floatfix]{revtex4-1}
\usepackage[svgnames]{xcolor}
\definecolor{darkred}{rgb}{0.6, 0, 0}
\definecolor{darkgreen}{rgb}{0, 0.5, 0}
\usepackage[colorlinks=true,urlcolor = teal,linkcolor = darkgreen, citecolor = violet,bookmarks=false]{hyperref}
\usepackage{amsfonts,amsmath,amssymb,amsthm}
\usepackage{graphicx}
\usepackage{dsfont}
\usepackage[normalem]{ulem}
\usepackage{enumerate}

\usepackage{tikz,tikz-3dplot}
\usetikzlibrary{arrows,patterns,positioning,shadows}
\usepackage{pgfplots}
\pgfplotsset{
grid style={dashed,gray!20}
}
\pgfplotscreateplotcyclelist{my black white}{%
solid, every mark/.append style={solid, fill=gray}, mark=diamond*\\%
dotted, every mark/.append style={solid, fill=gray}, mark=square*\\%
}

\usepackage{tcolorbox}
\usepackage{tabularx}
\usepackage{array}
\usepackage{colortbl}
\tcbuselibrary{skins}
\newcolumntype{C}{>{\centering\arraybackslash}X}
\newcolumntype{Y}{>{\raggedleft\arraybackslash}X}

\swapnumbers
\newtheorem{thm}{Theorem}[section]

\theoremstyle{plain} % just in case the style had changed
\newcommand{\thistheoremname}{}
\newtheorem{genericthm}[thm]{\thistheoremname}

\tcbset{tab/.style={enhanced,fonttitle=\bfseries,fontupper=\small\sffamily,
colback=green!5!white,colframe=teal,colbacktitle=red!40!white,coltitle=black,center title}}

\def\ii{{\rm i}}
\newcommand{\dd}{{\rm d}}

\def\expect#1{\langle#1\rangle}

\def\ol#1{\bar{#1}}

\newcommand{\rH}{\mathrm{H}}
\newcommand{\rX}{\mathrm{X}}

\newcommand{\U}{\mathrm{U}}

\newcommand{\SU}{\mathrm{SU}}

\begin{document}

\title{Undular diffusion in nonlinear sigma models}

\author{\v{Z}iga Krajnik}
\affiliation{Faculty for Mathematics and Physics,
University of Ljubljana, Jadranska ulica 19, 1000 Ljubljana, Slovenia}

\author{Enej Ilievski}
\affiliation{Faculty for Mathematics and Physics,
University of Ljubljana, Jadranska ulica 19, 1000 Ljubljana, Slovenia}

\author{Toma\v{z} Prosen}
\affiliation{Faculty for Mathematics and Physics,
University of Ljubljana, Jadranska ulica 19, 1000 Ljubljana, Slovenia}

\date{\today}

%%%%%%%%%%%%%%%%%%%%%%%%%%%%%%%%%%%%%%%%%
%%%%%%%%%%%%%%%%%%%%%%%%%%%%%%%%%%%%%%%%%
\begin{abstract}
We discuss general features of charge transport in non-relativistic classical field theories invariant under non-abelian unitary
Lie groups by examining the full structure of two-point dynamical correlation functions in grand-canonical ensembles at finite
charge densities (polarized ensembles). Upon explicit breaking of non-abelian symmetry, two distinct transport laws 
characterized by dynamical exponent $z=2$ arise. While in the unbroken symmetry sector the Cartan fields exhibit normal diffusion,
the transversal sectors governed by the nonlinear analogue of Goldstone modes disclose an unconventional law of diffusion
characterized by a complex diffusion constant and undulating patterns in the spatiotemporal correlation profiles.
In the limit of strong polarization, one retrieves the imaginary-time diffusion for uncoupled linear Goldstone modes,
whereas for weak polarizations the imaginary component of the diffusion constant becomes small.
In models of higher rank symmetry, we prove absence of dynamical correlations among distinct transversal sectors.
\end{abstract}

\pacs{02.30.Ik,05.70.Ln,75.10.Jm}

\maketitle
%\paragraph*{\bf Introduction.} 

Field theories provide one of the most invaluable tools in theoretical physics, with countless applications across a wide range
of disciplines. One of the most renowned and best studied examples are nonlinear sigma models (NLSMs) \cite{ZZ78,ALV78,Mikhailov82,PW83,Haldane83} and extensions thereof such as
Wess--Zumino--Witten models \cite{WZ71,Witten83,Witten84}, representing field theories of interacting fields on curved manifolds
that transform as representations of non-abelian symmetry groups.
Although sigma models have played a pivotal role in the studies of Yang--Mills theories and gauge-gravity dualities 
\cite{MZ03,Kazakov2004,Stefanski04}, renormalization group flows \cite{BZ76,Symanzik83}, topological QFTs \cite{Haldane83,Witten88}
and quantum criticality \cite{Sachdev99,Sachdev_book}, their dynamical properties remain poorly understood,
especially so in thermal equilibrium. One notable exception is the quantum ${\rm O}(3)$ NLSM in two space-time dimensions,
a prominent example of an integrable quantum field theory (QFT) \cite{ZZ78,Mikhailov82,ORW87,ZZ92} which has attracted a
considerable amount of attention in the context of low-temperature magnetization transport in Haldane antiferromagnets~\cite{Haldane83} 
(see \cite{SD97,DS98,Fujimoto99,Konik2003}), recently revisited in \cite{NMKI19}. Despite many efforts in the domain of quantum field 
theories \cite{EK09,Sachdev_book}, and recently even in classical isotropic magnets \cite{Gamayun2019,NMKI20,KP20,MatrixModels}, a 
comprehensive understanding of dynamical properties of NLSMs in thermal equilibrium is still lacking.

Our study is motivated by the following fundamental question: consider $G$-invariant NLSMs with coset spaces $\mathcal{M}=G/H$
as their target manifolds, where isometry group $G$ is a non-abelian simple Lie group and isotropy subgroup $H\subset G$
identified with stability group of a continuously degenerate vacuum state. As a consequence of $G$-invariance, the system possess
conserved Noether currents. The goal is a general classification of transport laws in thermal equilibrium states, 
irrespectively of the coset structure, Lorentz invariance, dimensionality, and integrability. In this Letter, we make a key progress
in this direction and classify dynamical two-point correlation functions in equilibrium states at generic values of background charge 
densities for a family of classical non-integrable NLSM in two space-time dimensions.

There is a widespread belief that \emph{ergodic} (chaotic) interacting systems governed by reversible microscopic dynamical laws 
exhibit normal diffusion, epitomized by the celebrated Fick's second law $\phi_{t}=D\,\phi_{xx}$ \cite{Fick} (unless several conservation laws are nontrivially coupled in 
which case nonlinear fluctuating hydrodynamics \cite{herbert} predicts a plethora of superdiffusive scaling laws \cite{slava}).
Here $\phi$ is a real scalar field whose spatial integral is conserved under time evolution, $(\dd/\dd t)\int\!\dd x\,\phi = 0$.
More generally, one speaks of \emph{normal diffusion} (in thermal equilibrium) when asymptotic dynamical structure factors,
reading $\expect{\phi(x,t)\phi(0,0)} \simeq t^{-1/z}f_{\rm G}\big((\lambda\,t)^{-1/z}x\big)$, are characterized by (i) dynamical exponent $z=2$ and (ii) \emph{Gaussian}
stationary scaling profile $f_{\rm G}(\zeta)=\exp{(-\zeta^{2})}$, parametrized by a {\em real} state-dependent (diffusion)
constant $D=\lambda/4$. In what follows, we shall explain how in systems with non-abelian continuous symmetries conserved Noether 
charges from the symmetry-broken sectors evade the conventional paradigm of normal diffusion.

%\paragraph*{\bf Introducing undular diffusion.}
\paragraph*{\bf Undular diffusion at a glance.}
The theme of this paper is an anomalous type of diffusion law we dub as `undular diffusion'.
To set the stage, we would first like to offer some basic intuition behind this notion. To this end, we consider a classical
isotropic ferromagnet. The vacuum (minimum energy configuration) corresponds to all the spins aligning in the same direction, taking the role of a local 
order parameter. The order parameter always picks a random polarization direction (a unit vector on a $2$-sphere), while the rotational 
symmetry of the model implies that the vacuum state is continuously degenerate and the symmetry is said to
be \emph{spontaneously} broken. It is widely known that a spontaneous breaking of continuous symmetry is accompanied by
soft Nambu--Goldstone modes; in ferromagents specifically, these are quadratically dispersing magnons which resolve small fluctuations 
about the symmetry-broken ferromagnetic vacuum.

\begin{figure*}[t]
\centering
\includegraphics[width=0.9\textwidth]{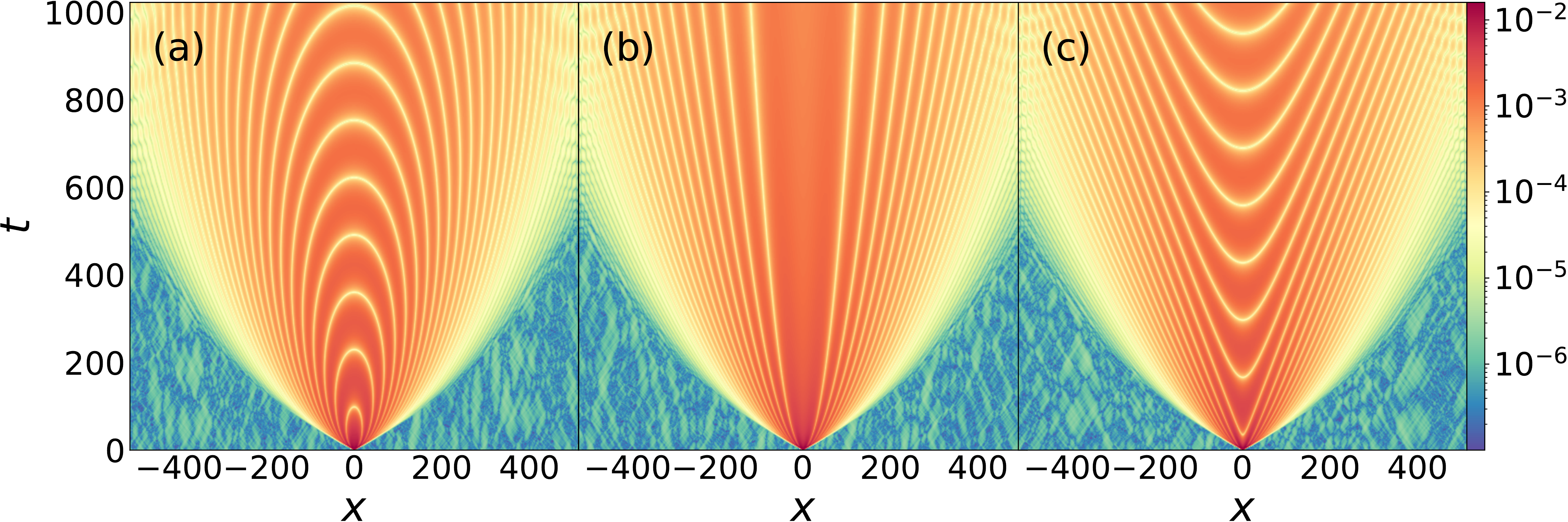}
\caption{Dynamical correlation functions in the transversal sector, computed in a non-integrable space-time lattice discretization
of $S^2$ Landau--Lifshitz field theory \cite{suppmat}, immersed in a longitudinal magnetic field of magnitude $b$ (pointing in the $z$-
direction). We display $\expect{S^{x}(x,t)S^{x}(0,0)}_{\mu}$ evaluated in a grand-canonical ensemble at infinite temperature and 
chemical potential $\mu=5$ ($\expect{S^z} \approx 0.8$), shown in absolute value (time-step $\tau=1$, length $L=1024$,
average over $7.5\cdot 10^{5}$ iterations). Three types of dynamical patterns can be discerned:
(a) elliptic regime ($b=-6\cdot 10^{-3}$), (b) parabolic regime of undular diffusion without a field ($b=0$), and (c) hyperbolic 
regime ($b=6\cdot 10^{-3}$). The characteristic curves resemble conic sections associated with linear Goldstone modes.}
\label{fig:panels}
\end{figure*}

Suppose we would like to understand transport properties of an isotropic ferromagnet at finite temperature. Invariance under continuous 
rotational symmetry implies that all the components of magnetization are globally conserved under time evolution. Transport properties 
of the model are most commonly extracted from the late-time relaxation of temporal correlation functions among distinct components.
In canonical Gibbs states, which respect the full rotational symmetry of the model there is no distinction between magnetization components.
By invoking standard hydrodynamic arguments (based on gradient expansion of local conserved currents), one expects to find normal 
diffusion governed by the aforementioned Fick's law.

Consider now the grand-canonical ensemble where rotational symmetry is \emph{explicitly} broken by inclusion of chemical potentials:
one polarization direction (and thereby magnetization component) becomes distinguished, while the remaining two components are 
proclaimed as transversal. The question is whether such a symmetry breaking scenario `at finite density' has any effect on
transport properties. One may indeed expect the difference to show up in the transversal sector; it is evident that in the limit of 
strong polarization, where thermal fluctuations are dominated by fluctuations near the ferromagnetic vacuum, one should recover 
precessional motion governed by the spectrum of Goldstone modes, which one can interpret as a diffusion process in imaginary time. Accordingly, it is natural to anticipate that at any intermediate density (i.e. finite chemical potential) the diffusive relaxation of 
transversal correlators acquires an extra imaginary component, combining into a single `complex Goldstone mode'.
This is precisely what happens, as shown in the remainder of the Letter.

\paragraph*{\bf Minimal example.}
We proceed by detailing out the `minimal model' of undular diffusion: the classical Landau-Lifhsitz field theory
\cite{Takhtajan1977,Faddeev1987} (using subscripts to designate partial derivatives)
\begin{equation}
{\bf S}_{t}={\bf S}\times {\bf S}_{xx}+{\bf S}\times {\bf B},
\label{eqn:LL}
\end{equation}
written in terms of the unit vector (spin) field ${\bf S}\equiv(S^{\rm x},S^{\rm y},S^{\rm z})^{\rm T}$ taking values on
a $2$-sphere, ${\bf S} \cdot {\bf S} = 1$. We have also included an external magnetic field ${\bf B}=b\,{\bf \hat{e}}_{\rm z}$ 
aligned with the vacuum polarization axis ${\bf \hat{e}}_{\rm z}=(0,0,1)^{\rm T}$ to study also `dynamical' breaking of symmetry.
%The dynamics (\ref{eqn:LL}) is generated by the Hamiltonian
%$H_{S^2} + H_{\bf B} = \frac{1}{2}\int_{\mathbb{R}}\! \dd x\,{\bf S}^{2}_{x} + \int_{\mathbb{R}} \dd x \, {\bf S} \cdot  {\bf B}$.

To study dynamics in the symmetry-broken states, we introduce transversal complex fields $S^{\pm}= S^{x} \pm \ii S^{y}$.
In Fig.~\ref{fig:panels} we display the dynamical correlator $\frac{1}{2}\textrm{Re}\expect{S^{+}(x,t)S^{-}(0,0)}_{\mu}$, averaged
with respect to the invariant grand-canonical Gibbs state at `infinite temperature' and chemical potential $\mu$ with a local probability density
$\varrho^{(1)}_\mu({\bf S})=(\pi \sinh{(\mu)}/\mu)^{-1}\exp{\big(\mu(1-2S^{z})\big)}$. To avoid special features related to integrability
of Eq.~\eqref{eqn:LL}, we performed our simulations on a {\em non-integrable} lattice discretization, see \cite{suppmat}.

In the absence of an external magnetic field we encounter undular diffusion, manifesting itself in the form of a spatially undulating 
correlation function with a characteristic diffusive (parabolic) pattern displayed in Fig.~\hyperref[fig:panels]{1b}.
When the rotational symmetry of the model is `dynamically broken' with an external positive (negative) magnetic field field,
$\int_{\mathbb{R}} S^{\pm}(x, t) \dd x = e^{\mp\ii bt}\int_{\mathbb{R}} S^{\pm}(x, 0) \dd x$, we observe
hyperbolic (elliptic) characteristics, as shown in Figs.~\hyperref[fig:panels]{1a},\hyperref[fig:panels]{1c}.

\begin{figure*}
%\centering
%%%%%%%%%%%%%%%%%%% PROFILES %%%%%%%%%%%%%%%%%%%%
\begin{minipage}[t]{0.64\linewidth}
\raggedleft
\includegraphics[width=\textwidth]{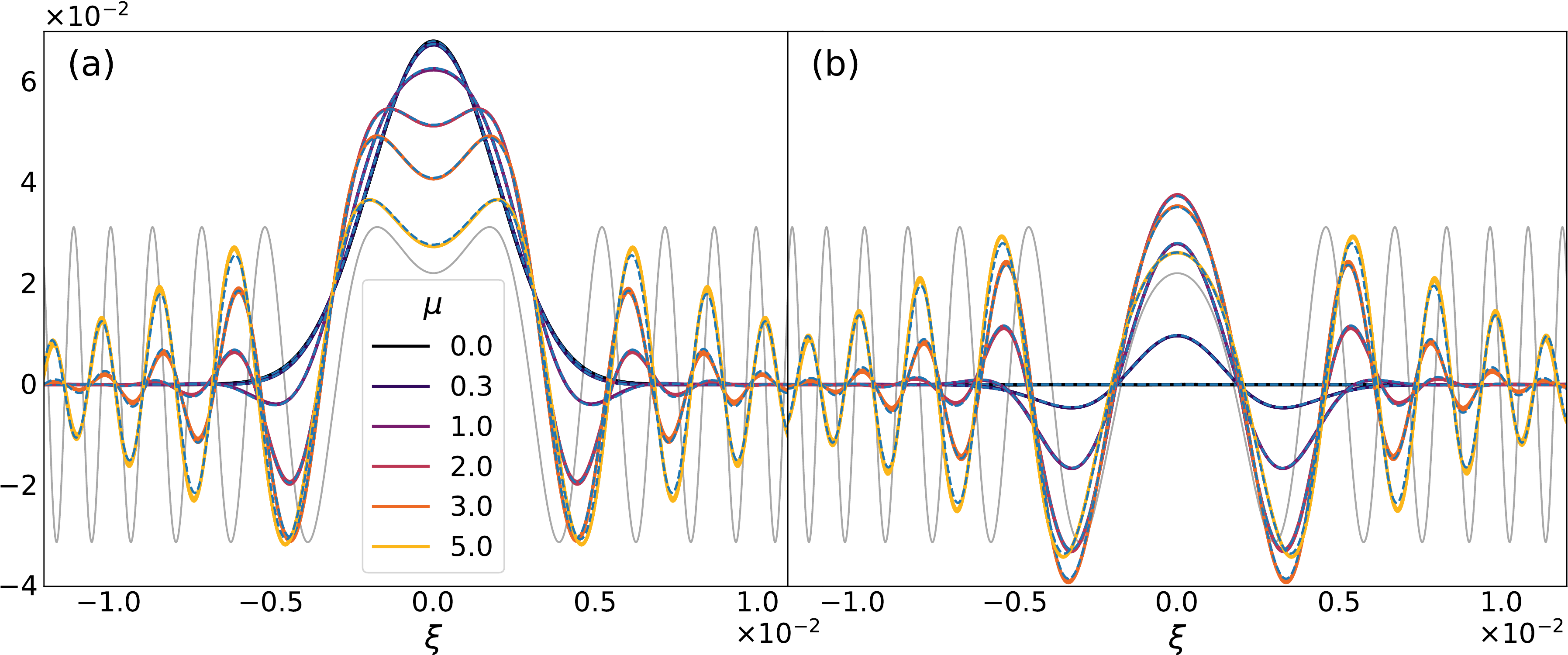}
\label{fig:profiles}
\end{minipage}
\begin{minipage}[t]{0.35\linewidth}
\raggedleft
\includegraphics[width=\textwidth]{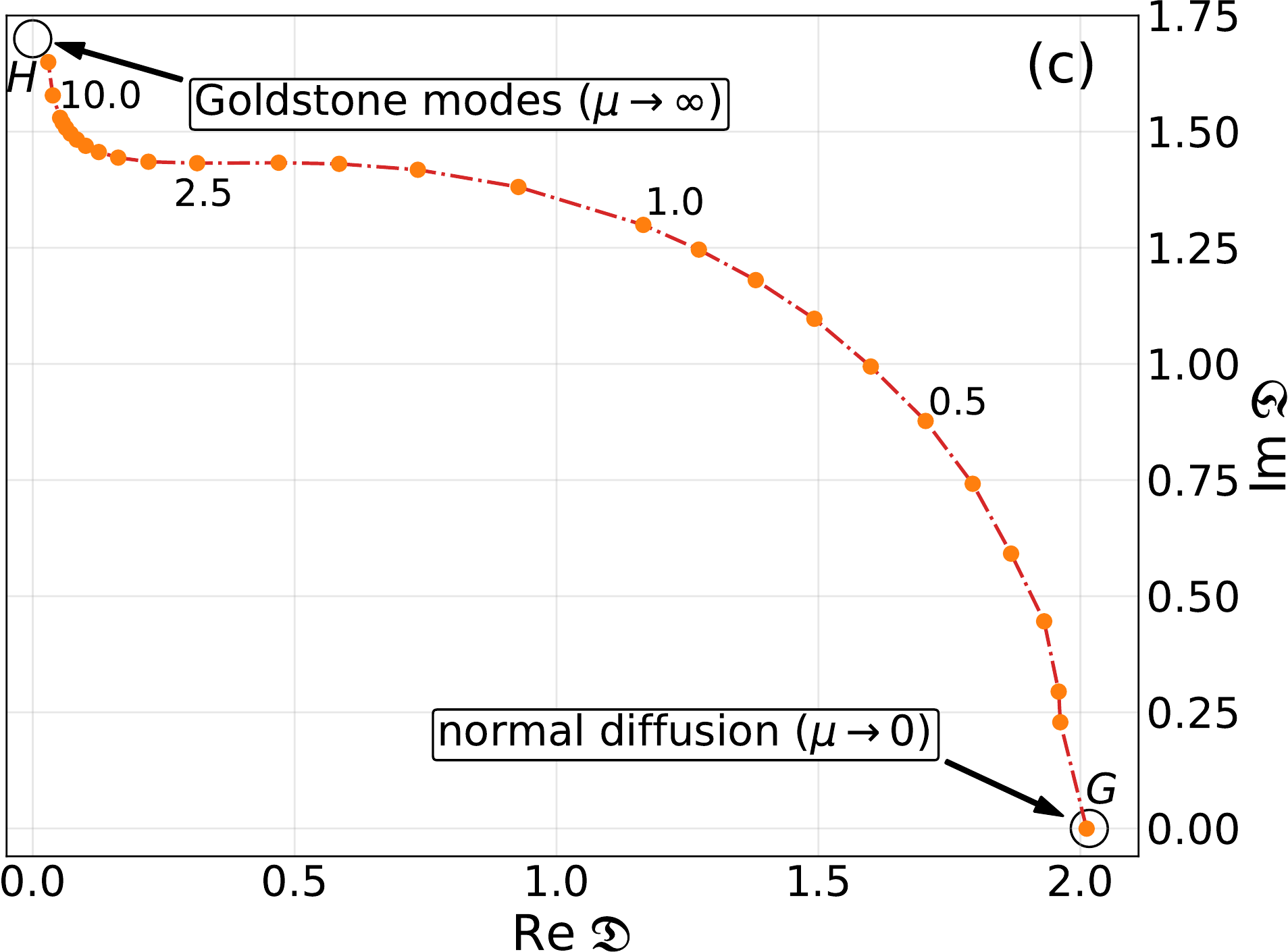}
\label{fig:diffusion_constant}
\end{minipage}
\caption{Stationary asymptotic profiles of the transversal Goldstone mode in the non-integrable space-time lattice discretization
of  the $S^2$ Landau--Lifshitz field theory \cite{suppmat} without a field ($b=0$), displaying (a) real and (b) imaginary components of
$\lim_{t\to \infty}|t|^{1/2}\expect{S^{+}(x,t)S^{-}(0,0)}_{\mu}/2$ as a function of the scaled variable $\xi=t^{-1/2}x$ and $\U(1)$ 
chemical potential $\mu$ (using same parameters as in Fig.~\hyperref[fig:panels]{1b}). Gray curves mark the prediction of the
linear theory, cf. Eq.~\eqref{eqn:linear} (in arbitrary units). Dashed blue lines show best two-parameter fits to
Eq.~\eqref{scaling_function}.({c}) Dependence of real and imaginary components of complex diffusion constant $\mathfrak{D}$
on chemical potential $\mu$. Symmetry points $H$ and $G$ designate the $U(1)$-invariant vacuum and $SU(2)$-invariant equilibrium 
measure, respectively (red dashed line is a guide to the eye).}
\label{fig:profiles}
\end{figure*}

The origin of the observed patterns is best explained by inspecting the vicinity of the ferromagentic vacuum ($|S^{z}| \rightarrow 1$)
where the equation of motion in the transversal sector reduces to a \emph{linear} theory \cite{suppmat}
\begin{equation}
\Big[\ii \partial_{t}\pm \big(\partial^{2}_{x} - b\big)\Big] S^{\pm}(x,t) = 0.
\label{eqn:linear}
\end{equation}
Its Green's function $\mathcal{G}_{\pm}(k)=\exp{\left[\mp\,\omega^{b}_{\rm mag}(k)t\right]}$
describes magnons with a gapped quadratic (i.e. type-II) dispersion law
\begin{equation}
\omega^{b}_{\rm mag}(k)=\mathfrak{D}_{\infty}k^{2}+\ii b,
\label{eqn:dispersion}
\end{equation}
written as an imaginary-time diffusion with an imaginary diffusion constant
$\mathfrak{D}_{\infty}\equiv \mathfrak{D}(\mu \to \infty)=\ii$. Characteristics associated with Eq.~\eqref{eqn:linear} are conic 
sections. Remarkably, their presence remains visible even in the nonlinear dynamic away from the vacuum (i.e. at general values
of $\mu$), as shown in Fig.~\ref{fig:panels}.

%yielding $S^{\pm}(x,t)=\int \dd x'\,\mathcal{G}_{\pm}(x-x', t)S^{\pm}(x',0),$
%with convolution kernels of the form
%\begin{equation}
%\mathcal{G}_{\pm}(x,t) =\big(4\pi |t|\big)^{-1/2}\exp{\left [\mp\, \ii\left( b|t|-\frac{x^{2}}{4|t|}+ \frac{\pi}{4} \right) \right]}.
%\label{Green}
%\end{equation}
%Zeros of ${\rm Re}\,\mathcal{G}_{\pm}(x,t)$ (resp. ${\rm Im}\,\mathcal{G}_{\pm}(x,t)$) lie along conic sections (which degenerate into parabolae at $b=0$)
%\begin{equation}
%4b(t^{2}-2t^{b}_{\rm n}t)=x^{2}, \qquad
%t^{b}_{\rm n}  \equiv \frac{\pi}{2b}({\rm n}+1/4),
%\end{equation} with integer (resp. half-integer) ${\rm n}$ in agreement with the three regimes observed in Fig.~\ref{fig:panels}.

In Fig.~\ref{fig:profiles} we display the numerically computed stationary profiles for the transversal dynamical correlator
(without the field, $b=0$) depending on chemical potential $\mu$. When approaching the vacuum (i.e. at large $\mu$),
the profiles converge towards the prediction of the linear theory \cite{suppmat} (grey curves in Fig.~\ref{fig:profiles}).
In the opposite regime, $\mu \to 0$, the profiles smoothen out into a Gaussian. In Fig.~\hyperref[fig:profiles]{2c}
we extract the complex diffusion constant $\mathfrak{D}(\mu)$ by fitting the scaling function (given below
in Eq.~\ref{scaling_function}).

The above phenomenology offers the following suggestive interpretation: at finite spin density ($\mu\neq 0$), the late-time relaxation of nonlinearly
evolving fields from the symmetry-broken sector is governed by an unconventional Goldstone mode which has acquired
an extra diffusive component, characterized by a single hydrodynamic generalized Fick's law of diffusion with
a \emph{complex} diffusion constant $\mathfrak{D}(\mu)$.

To conclude, we remark that that two-point correlations $\expect{S^{\rm z}(x,t)S^{\pm}(0,0)}_{\mu}$ and
$\expect{S^{\pm}(x, t)S^{\pm}(0, 0)}_{\mu}$ both trivially vanish as consequence of the residual $U(1)$ symmetry
about ${\bf e}_{\rm z}$. The only remaining non-zero correlator allowed by symmetry is therefore the longitudinal one,
$\expect{S^{\rm z}(x, t)S^{\rm z}(0, 0)}_{\mu}$, which, as expected, undergoes normal diffusion with real diffusion constant.

\paragraph*{\bf Symmetry of higher rank.}
It is natural to ask if any new feature can arise in models exhibiting symmetries of higher rank.
Our next aim thus is to classify the dynamical two-point correlation functions among the Noether charges of a class of models
invariant under non-abelian groups of higher rank, comprising multiple Nambu--Goldstone modes in their spectrum.
We mainly wish to discern whether enhanced symmetry can affect dynamics in the symmetry-broken sector due to interaction
among distinct transversal modes.

Here we shall consider the simplest class of (non-relativistic) continuous ferromagnets invariant under the action of
unitary Lie groups $\SU(n+1)$, whose target spaces are complex projective manifolds $\mathcal{M}_{n}\equiv \mathbb{CP}^{n}$.
The latter are naturally parametrized by complex fields $z^{a}(x,t)$ (alongside their conjugate counterparts $\ol{z}^{a}$),
and for compactness we introduce the vector of affine coordinates ${\bf z}\equiv (z_{1},\ldots,z_{n})^{\rm T}$ on $\mathcal{M}_{n}$.
As a starting point, we consider the most general effective Lagrangian invariant under $\SU(n+1)$ in the form
\begin{equation}
\mathcal{L}_{\rm eff} \simeq \mathcal{L}_{\rm WZ} - \mathcal{L}^{(2)}_{\mathbb{C}\mathbb{P}^{n}} + {\tt higher\;order\;terms},
\label{eqn:effective_Lagrangian}
\end{equation}
where $\mathcal{L}^{(2)}_{\mathbb{C}\mathbb{P}^{n}}\equiv \sum_{a,b=1}^{n}\eta_{ab}\ol{z}^{a}_{x}z^{b}_{x}$
is the second-order term in gradient expansion parametrized by the unique $G$-invariant Riemann (Fubini--Study) metric
on $\mathbb{CP}^{n}$, reading explicitly
$\eta_{ab}=((1+{\bf z}^{\dagger}{\bf z})\delta_{ab}-\ol{z}^{a}z^{b})/(1+{\bf z}^{\dagger}{\bf z})^{2}$, and
$\mathcal{L}_{\rm WZ}=\ii(1+{\bf z}^{\dagger}{\bf z})^{-1} ({\bf z}^{\dagger}{\bf z}_{t}-{\bf z}^{\dagger}_{t}{\bf z})$
denotes the Wess--Zumino geometric term.

To simplify our analysis, we shall discard all the higher-order terms in Eq.~\eqref{eqn:effective_Lagrangian}.
This way, we end up with non-relativistic classical sigma models on cosets $\mathbb{C}\mathbb{P}^{n} = G/H$,
with isotropy subgroup $H = SU(n)\times U(1)$ leaving the ferromagnetic vacuum intact (modulo a phase).
Matrix-valued fields $M(x,t)$ on $\mathcal{M}_{n}$ are unitary matrices subjected to a nonlinear constraint
$M^{2}=\mathds{1}$, in terms of which the Hamiltonian reads simply
$H_{\mathbb{C}\mathbb{P}^{n}}=2\int \dd x\,\mathcal{L}^{(2)}_{\mathbb{C}\mathbb{P}^{n}}
=\frac{1}{4}\int\! \dd x\,{\rm Tr}\big(M^{2}_{x}\big)$. The equation of motion is given by a nonlinear PDEs
(Landau--Lifshitz field theories of higher-rank) \cite{KP20}
\begin{equation}
M_{t} = \frac{1}{2\ii}[M,M_{xx}]+\ii[B,M].
\label{eqn:EOM}
\end{equation}
where we have simultaneously adjoining the external field $H_{B}=\int\!\dd x\,{\rm Tr}(B\,M)$ which induces
dynamical breaking of conservation laws associated to $G$.

Longitudinal and transversal fields can be inferred with respect to the Cartan-Weyl basis of Lie algebra
$\mathfrak{g}=\mathfrak{su}(n+1)$ (see e.g. \cite{Humphreys_book,Hall_book}, and \cite{suppmat} for details).
Weyl generators, which are indexed by root vectors spanning the root lattice $\Delta$ of $\mathfrak{g}$, are assigned complex Weyl 
fields $\phi^{\pm \alpha}$. To every Cartan generator we associate a real longitudinal field $\phi^{i}$ and formally assign to it a 
`zero root' forming a set $\Delta_{0}$. To obtain the $\phi$-field, one simply traces the corresponding generator times the matrix $M \in \mathcal{M}_{n}$.	

We proceed by introducing grand-canonical Gibbs states, including generic chemical potentials coupling to the Cartan charges
$\mathcal{Q}^{i}=\int \dd x\,\phi^{i}(x)$. In such a state, the original symmetry $G=\SU(n+1)$ gets lowered down to the residual symmetry 
of its maximal abelian subgroup $\mathrm{T}=\U(1)^{\times n}$. There are thus $n$ `unbroken' longitudinal fields $\phi^{i}$ associated to the Cartan generators. On the other hand, the symmetry-broken sector comprises of
$n_{t}=\tfrac{1}{2}{\rm dim}(G/{\rm T})=\tfrac{1}{2}n(n+1)$ pairs
of canonically-conjugate complex `transversal' modes $\phi^{\pm \alpha}$.
To define a stationary measure invariant under ${\rm T}$, we introduce the diagonal `torus Hamiltonian'
$\rH_{{\boldsymbol{\mu}}}=-\tfrac{1}{2}{\rm diag}(\mu_{0},\mu_{1},\ldots,\mu_{n})$, parametrized by chemical potentials
$\mu_{i}\in \mathbb{R}$ (subjected to ${\rm Tr}\,\rH_{\boldsymbol{\mu}}=0$) and define an invariant normalized measure 
$\varrho^{(n)}_{\boldsymbol{\mu}}\dd \Omega^{(n)}$ ($\int_{\mathcal{M}_{n}}\dd \Omega^{(n)}\varrho^{(n)}_{\boldsymbol{\mu}}=1$),
with volume element $\dd \Omega^{(n)}$ on $\mathbb{CP}^{n}$ and density
\begin{equation}
\varrho^{(n)}_{\boldsymbol{\mu}}(M) = \frac{1}{\mathcal{Z}^{(n)}_{\boldsymbol{\mu}}}
\exp{\Big({\rm Tr}(\rH_{\boldsymbol{\mu}}M)\Big)},
\label{eqn:gc_measure}
\end{equation}
where $\mathcal{Z}^{(n)}_{\boldsymbol{\mu}} = \int_{\mathcal{M}_{n}} \dd \Omega^{(n)}\exp{\big({\rm Tr}(\rH_{\boldsymbol{\mu}}M)\big)}$ 
represents the partition function. One can think of Eq.~\eqref{eqn:gc_measure} as the grand-canonical Gibbs measure at infinite
temperature (known in symplectic geometry as an equivariant measure). Further details can be found in \cite{suppmat}.\\

By direct analogy to the previous basic case of $\mathbb{CP}^{1}\cong S^{2}$,
the mixed correlators $\expect{\phi^{i}(x,t)\phi^{\pm \alpha}(0,0)}_{\boldsymbol{\mu}}$ and 
paired intrasectoral correlators $\expect{\phi^{\pm \alpha}(x,t)\phi^{\pm \alpha}(0,0)}_{\boldsymbol{\mu}}$
once again vanish as a direct corollary of the ${\rm T}$-invariance of the measure \eqref{eqn:gc_measure}.
Indeed, this statement remains valid even in Gibbs states at any inverse temperature $\beta$.

The new ingredient now is that models of higher rank possess additional intersectoral correlations among distinct transversal (Weyl) 
fields. A starting point for their analysis is the following `neutrality selection rule' for equal-time $N$-point correlators
\begin{equation}
\sum_{j\in\{1\ldots N\}}^{\sigma_{j}\not\in\Delta_{0}}\sigma_{j} 
\neq {\bf 0} \quad \Rightarrow \quad
\langle \phi_{\ell_1}^{\sigma_1} \phi_{\ell_2}^{\sigma_2} \dots \phi_{\ell_N}^{\sigma_N} \rangle_{\beta, \pmb\mu} = 0,
\label{neutrality_rule}
\end{equation}
which, in conjunction with the commutation relations, implies (see \cite{suppmat} for proofs) the `kinematic' decoupling
of transversal modes into subsectors, that is
\begin{equation}
\expect{\phi^{\pm \alpha}(x,t)\phi^{\gamma \nparallel \alpha}(0,0)}_{\beta, {\boldsymbol{\mu}}} = 0. 
\end{equation}
Consequently, the only dynamical two-point correlation functions allowed by symmetry are, besides
the longitudinal $\expect{\phi^{i}(x,t)\phi^{j}(0,0)}_{\boldsymbol{\mu}}$, the {\em intrasectoral} correlations
$\expect{\phi^{\pm \alpha}(x,t)\phi^{\mp \alpha}(0,0)}_{\boldsymbol{\mu}}$.

Numerical analysis of asymptotic stationary profiles within each transversal `$\alpha$-sector' shows that asymptotic dynamical structure factors are accurately captured by scaling profiles of undular diffusion
\begin{equation}
\expect{\phi^{\alpha}(x,t)\phi^{-\alpha}(0,0)}_{\boldsymbol{\mu}}
= \frac{\chi_{\alpha,-\alpha}}{(4\pi\mathfrak{D}_{\alpha}|t|)^{1/2}}
e^{-x^{2}/\left(4\mathfrak{D}_{\alpha}|t|\right) },
\label{scaling_function}
\end{equation}
characterized by a \emph{complex} diffusion constant $\mathfrak{D}_{\alpha} (\pmb \mu) $ which recombines the effects of
relaxation and precessional motion into a {\em single hydrodynamic mode}. In the strong-polarization limit, we recover the frequency
of the (linear) Goldstone modes, $\lim_{|\pmb\mu|\to \infty}\mathfrak{D}_{\alpha}(\pmb \mu) = \ii\,\omega_{\alpha} (\langle \phi^j \rangle_{\textrm{vac}})$, whereas in the opposite regime of weak polarization $\lim_{|\pmb \mu| \rightarrow 0}\mathfrak{D}_{\alpha} (\pmb \mu) = D_{\alpha}  \in \mathbb{R}$ \footnote{Modulo possible numerically unresolved logarithmic corrections \cite{NMKI20}.}.

%
%\paragraph*{\bf Lattice discretizations.}
%The price for disregarding higher-order terms in the gradient expansion \eqref{eqn:effective_Lagrangian} is integrability of
%Eq.~\eqref{eqn:EOM}. To exclude non-generic effects, we purposefully destroyed integrability in our numerical simulations. A simple way to achieve this
%is via (generically) \emph{non-integrable} lattice discretizations. Denoting by $M_{\ell}$ a matrix variable at site $\ell$,
%we consider a local `precession law' of the form
%\begin{equation}
%\frac{\dd M_{\ell}}{\dd t} = -\ii\Big[M_{\ell},\mathcal{F}(\mathbb{M}_{\ell-1,\ell})+\mathcal{F}(\mathbb{M}_{\ell,\ell+1})\Big]
%+\ii \Big[B,M_{\ell}\Big],
%\label{eqn:nonint}
%\end{equation}
%with $\mathbb{M}_{\ell,\ell+1}\equiv M_{\ell}+M_{\ell+1}$ and some appropriate analytic function $\mathcal{F}$.
%
%Efficiency of numerical simulations can be appreciably improved (without affecting transport properties of the Noether charges,
%see e.g. \cite{VZP18,LZP19,KP20}, by further breaking invariance under time translation. We achieved this via non-integrable
%Trotterization \cite{Trotter59} of Eq.~\eqref{eqn:nonint}, cf. \footnote{Supplemental Material associated with this manuscript}.

\paragraph*{\bf Summary.}
A succinct summary of our results is given in Table \ref{tab:correlators}.
%\begin{figure}[htb]
%\centering
%\includegraphics{profiles.pdf}
%\caption{Stationary asymptotic profiles of the transversal Goldstone mode in the non-integrable Trotterization of the $S^{2}$ NLSM, 
%displaying (a) real and (b) imaginary components of $\lim_{t\to \infty}|t|^{1/2}\expect{S^{+}(x,t)S^{-}(0,0)}_{\mu}/2$ as a function
%of the scaled variable $\xi=t^{-1/2}x$ and $\U(1)$ chemical potential $\mu$ (using same parameters as in Fig.~\ref{fig:panels}).
%Gray curves mark the prediction of the linear theory (in arbitrary units).}
%\label{fig:profiles}
%\end{figure}
Dynamical (connected) two-point correlations functions can be grouped into \emph{three} classes:
\begin{enumerate}[(I)]
\item \emph{longitudinal} correlations $\expect{\phi^{i}(x,t)\phi^{j}(0,0)}_{{\boldsymbol{\mu}}}$,
with dynamical exponent $z=2$ and Gaussian asymptotic profiles \footnote{This is based on numerical analysis of stationary asymptotic profiles in non-integrable space-time discretizations of $\mathbb{C}\mathbb{P}^{n}$ sigma models for $n\in\{1,2\}$.},
\item  \emph{transversal} `$\alpha$-sectors'
$\expect{\phi^{\pm \alpha}(x,t)\phi^{\mp \alpha}(0,0)}_{{\boldsymbol{\mu}}}$, with dynamical exponent $z=2$
and \emph{undulating} asymptotic stationary profiles (examplified for $n=1$ in Fig.~\ref{fig:profiles}),
\item (i) vanishing mixed and transversal correlations
$\expect{\phi^{i}(x,t)\phi^{\pm \alpha}(0,0)}_{{\boldsymbol{\mu}}}=
\expect{\phi^{\pm \alpha}(x,t)\phi^{\pm \alpha}(0,0)}_{{\boldsymbol{\mu}}}=0$,\\
and (ii) vanishing \emph{intersectoral} correlations 
$\expect{\phi^{\pm \alpha}(x,t)\phi^{\gamma \nparallel \alpha}(0,0)}_{{\boldsymbol{\mu}}}=0$.
\end{enumerate}

\begin{table}[b]
\centering
% leave the first two column's width adjustable
\begin{tcolorbox}[tab,tabularx*={\arrayrulewidth0.20mm}{c|c|C}]
{\tt sector} & {\tt correlators} & {\tt transport} \\
\hline\hline
{\rm longitudinal} &  $\expect{\phi^{i}(x,t)\phi^{j}(0,0)}_{\boldsymbol{\mu}}$ & {\rm normal diffusion} \\
\hline
{\rm transversal} & $\expect{\phi^{\pm \alpha}(x,t)\phi^{\mp \alpha}(0,0)}_{\boldsymbol{\mu}}$ & \textcolor{blue}{\rm undular diffusion} \\
\hline
  & $\expect{\phi^{\pm \alpha}(x,t)\phi^{\pm \alpha}(0,0)}_{\boldsymbol{\mu}}$ &  \\
{\rm trivial} & $\expect{\phi^{i}(x,t)\phi^{\pm \alpha}(0,0)}_{\boldsymbol{\mu}}$ & {\rm no transport} \\
 & $\expect{\phi^{\pm \alpha}(x,t)\phi^{\gamma \nparallel \alpha}(0,0)}_{\boldsymbol{\mu}}$ & 
\end{tcolorbox}
\caption{Complete classification of dynamical two-point correlation functions among the Noether fields.
}
\label{tab:correlators}
\end{table}

Properties (I) and (II) have been established based on numerical  observations, while (III-i) is a direct corollary
of invariance under the torus subgroup ${\rm T}$. Property (III-ii) follows from the `neutrality rule' \eqref{neutrality_rule}.
Indeed, we believe (I)-(III) are generic properties of non-integrable Hamiltonian dynamics invariant under non-abelian compact Lie 
group $G$ with $G/H$-valued local degrees of freedom (order parameter), averaged with respect to a polarized ${\rm T}$-invariant 
ensemble. In effect, the listed properties likewise apply to dynamical two-point functions in grand-canonical Gibbs ensembles at finite 
temperature, which will experience an additional `smearing' effect across a lengthscale comparable to the thermal correlation 
length.

\paragraph*{\bf Conclusion.}

Focusing on a class of non-relativistic sigma models invariant under unitary Lie groups, we have investigated the structure of 
dynamical correlations among the Noether charges in an equilibrium state with broken continuous symmetry.
While longitudinal correlations among  the Cartan fields expectedly undergo normal diffusion, we found that dynamics in the transversal 
(symmetry-broken) sector is governed by unorthodox Goldstone modes that satisfy a complexified diffusion law, characterized by 
dynamical exponent $z=2$ and `complex Gaussian' profiles governed by a \emph{complex} diffusion constant, which we have suggestively named \emph{undular diffusion}. The phenomenon is present in a generic non-integrable (chaotic) dynamics and does not depend on the 
microscopic details of the model or particular lattice discretization.

The main lesson to draw is twofold: (A) the ubiquitous Fick's law of diffusion, believed to be a hallmark of chaotic reversible
many-body dynamics, can indeed be violated in systems that support type-II Goldstone modes, and (B) dynamical systems
invariant under non-abelian Lie group $G$ do not support any dynamical correlations among
the conserved Noether currents from different transversal $\mathfrak{su}(2)$ sectors in grand-canonical Gibbs equilibrium states.
In regards to (A), an alternative viewpoint is to argue that undular diffusion is an analytic prolongation of the Fick's law of 
diffusion into the complex plane.

We note that classical non-relativistic NLSMs that appear as the leading term of the gradient expansion of $G$-invariant 
dynamics on hermitian symmetric spaces, such as Eqs.~\eqref{eqn:EOM}, are commonly found to be integrable \cite{Fendley99,Fendley01}.
A salient feature of integrable dynamics (which can be accurately captured by generalized hydrodynamics \cite{PhysRevX.6.041065,PhysRevLett.117.207201}) are stable nonlinear modes (solitons), which render 
longitudinal correlators ballistic (quantified by finite charge Drude weights \cite{IN_Drude,DS17,IN17,Bulchandani2018})
with diffusive corrections \cite{DeNardis2018,DeNardis_SciPost,Gopalakrishnan2018,Medenjak19},
or even superdiffusive dynamics that takes place in unpolarized Gibbs states \cite{MarkoKPZ,PZ13},
recently examined in \cite{Ljubotina2017,Ilievski2018,GV2019,NMKI19,DupontMoore2019,Vir2020,KP20,MatrixModels,NGIV20}.
In performing numerical computation we have always employed appropriate lattice discretizations to ensure that
integrability is manifestly broken. We have nonetheless verified that even in integrable discretizations of $\mathbb{C}\mathbb{P}^{n}$ 
sigma models \eqref{eqn:EOM} \cite{KP20}, dynamics of transversal models associated with internal `precessional' degrees of freedom 
still display undular diffusive profiles.

On general grounds one can expect that the phenomenon survives quantization, i.e. to persist in quantum lattice ferromagnets invariant 
under non-abelian Lie groups (irrespectively of integrability), and to extend to higher space-time dimensions.

There are several interesting venues left to be explored, for instance: (i) develop a quantitative framework to access
asymptotic stationary profiles that characterize undular diffusion; (ii) extend the analysis to other symmetry groups and
coset spaces; (iii) infer the structure of transversal dynamical correlators also in relativistic sigma models,
both in the classical and quantum settings.

\paragraph*{\bf Acknowledgement.}
We thank B. Bertini, S. Grozdanov and M. \v{Z}nidari\v{c} for insightful remarks.
The work has been supported by ERC Advanced grant 694544 – OMNES and the program P1-0402 of Slovenian Research Agency.

%%%%%%%%%%%%%%%%%%%%%%%%%%%%%%%%%%%%%%%%%
%%%%%%%%%%%%%%%%%%%%%%%%%%%%%%%%%%%%%%%%%

\bibliography{Undular}

\clearpage

\onecolumngrid

\begin{center}
\textbf{{\Large Supplemental Material}}
\end{center}
\begin{center}
\textbf{{\large Undular diffusion in nonlinear sigma models}}
\end{center}

\section{Preliminaries}

We consider a Hamiltonian dynamical system invariant under a unitary Lie group $G=\SU(n+1)$ with matrix fields $M(x,t)$ taking
values on complex projective spaces $M \in \mathcal{M}_{n} = \mathbb{C}\mathbb{P}^{n}$. The local phase space has
a structure of a quotient space (coset) $\mathbb{C}\mathbb{P}^{n} \cong G/H$, where $H=\SU(n)\times \U(1) \subset G$ is
the stability subgroup of a vacuum value $\Sigma^{(n)}$, that is $h\,\Sigma^{(n)}\,h^{-1}=\Sigma^{(n)}$ for $h \in H$.
With no loss of generality, we can set the polarization to
\begin{equation}
\Sigma^{(n)} = \mathds{1} - 2\Psi_{0} \Psi_{0}^\dagger,\qquad \Psi_{0} \equiv (1,0,\ldots,0)^{\rm T}.
\end{equation}
Hermitian matrices $M \in \mathcal{M}_{n}$ are then given by adjoint $G$-orbits of $\Sigma^{(n)}$,
\begin{equation}
M = g\,\Sigma^{(n)}\,g^{-1},
\end{equation}
and are subjected to the nonlinear constraint
\begin{equation}
M^{2} = \mathds{1}.
\label{app:nonlin}
\end{equation}

\medskip

\paragraph*{\bf Cartan--Weyl basis.}
By virtue of Eq.~\eqref{app:nonlin}, the number of independent components (i.e. scalar fields) of $M(x,t)$ equals
$2n$, that is the real dimension of complex manifold $\mathcal{M}_{n}$. To exhibit the underlying algebraic structure,
it proves most natural to employ the Cartan--Weyl basis of $\mathfrak{g}=\mathfrak{su}(n+1)$. Let $\rH^{i}$,
with $i\in \Delta_{0}\equiv\{1,2,\ldots,n\}$, denote the generators of the maximal abelian (Cartan) subalgebra,
\begin{equation}
[\rH^{i}, \rH^{j}] = 0,
\label{app:C_label}
\end{equation}
and $\rX^{\pm \alpha}$ the Weyl generators assigned to root vectors $\pm \alpha \in \Delta_{\pm}$, spanning the root lattice
$\Delta = \Delta_{+}\cup \Delta_{-}$. The defining algebraic relations in the Cartan--Weyl basis read
\begin{align}
[\rH^{i}, \rX^{\pm\alpha}] &= \pm \alpha^{i} \rX^{\pm \alpha},\\
[\rX^\alpha, \rX^{-\alpha}] & = \sum_{i,j=1}^{n}\alpha^{i} (\kappa^{-1})_{ij} \rH^{j},\\
[\rX^{\alpha}, \rX^{\gamma \neq - \alpha}] &= C_{\alpha, \gamma} \rX^{\alpha + \gamma}.
\label{app:CW_algebra}
\end{align}
where $\kappa_{ij} = {\rm Tr}(\rH^{i}\rH^{j})$ are matrix elements of the Killing form $\kappa$ and
for Weyl generators we adopted normalization ${\rm Tr}(\rX^\alpha \rX^{-\alpha}) = 1$ (with generators in the fundamental
representation). Weyl generators in the fundamental representation of $\mathfrak{su}(n+1)$ thus read explicitly
\begin{equation}
\big(\rX^{\gamma}\big)_{a,b} = \delta_{a,k}\delta_{b,k'} \qquad {\rm for}\qquad \gamma\equiv(k,k') \in \Delta.
\label{app:Weyl_explicit}
\end{equation}

\medskip

\paragraph*{\bf Cartan and Weyl variables.}
For computational convenience, we shall avoid a field-theoretical description that necessitates path-integral techniques and instead
provide an explicit lattice formulation. We thereby consider a one-dimensional lattice of $L$ sites, attaching a local matrix variable
$M_{\ell}\in \mathcal{M}_{n}$ to every site $\ell \in [1,L]$.
The global phase space is accordingly just an $L$-fold Cartesian product of all local target spaces $\mathcal{M}_{n}$.
Local matrix variables admit the `Cartan decomposition'
\begin{equation}
M_{\ell} = \frac{n-1}{n+1} \mathds{1} + \sum_{j=1}^{n}\phi_{\ell}^{j}  \rH^{j}
+ \sum_{\pm \alpha \in \Delta_{\pm}}\phi_{\ell}^{\pm \alpha} \rX^{\mp \alpha}.
\label{app:decomposition}
\end{equation}
Their components, 
\begin{equation}
\phi^{i}_{\ell} = \sum_{j=1}^{n}(\kappa^{-1})_{ij} {\rm Tr}(M_{\ell} \rH^{j}), \qquad
\phi^{\pm \alpha}_{\ell} = {\rm Tr}(M_{\ell} \rX^{\pm \alpha}),
\end{equation}
shall be referred to as the Cartan and Weyl fields, respectively.

\medskip

\paragraph*{ \bf Darboux coordinates.}

Complex projective spaces $\mathbb{C}\mathbb{P}^{n}$ are examples of toric manifolds. This signifies that they are
diffeomorphic to a product of a real $n$-torus $\mathbb{T}^{n}$ and an $n$-dimensional polytope, the standard
simplex $\varDelta^{(n)} \subset \mathbb{R}^{n}$, $\mathbb{C}\mathbb{P}^{n} \cong \varDelta^{(n)}\times \mathbb{T}^{n}$ \cite{Vergne96}.

Below we outline an explicit construction of canonical (Darboux) coordinates on $\mathcal{M}_{n}$, which will greatly
facilitate subsequent analytic considerations. Using that elements $M \in \mathcal{M}_{n}$ of complex projective spaces
can be expressed in terms of rank-$1$ projectors, we write
\begin{equation}
M=\mathds{1}-2\Psi\Psi^\dagger,\qquad \Psi=(\psi^{0},\psi^{1},\ldots,\psi^{n})^{\rm T},
\qquad \Psi^\dagger \Psi = \sum_{i=0}^{n}|\psi^{i}|^{2}=1
\end{equation}
where $\Psi$ are complex homogeneous coordinates of $\mathcal{M}_{n}$.
These can be in turn parametrized in terms of `octant coordinates' \cite{Bengtsson_book}
\begin{equation}
\psi^{j}=\nu^{j}\exp{(\ii \varphi^{j})},\qquad \nu^{j}\in [0,1],\quad \varphi^{j}\in [0,2\pi).
\end{equation}
Canonical variables of $\mathcal{M}_{n}$ are provided by angle coordinates $\varphi^{j}$ of $\mathbb{T}^{n}$
and conjugate momenta $p^{j}\equiv (\nu^{j})^{2}\in [0,1]$ that span the momentum polytope -- standard $n$-simplex $\varDelta^{(n)}$.
In Darboux coordinates, elements of matrix variables $M\in \mathcal{M}_{n}$ are therefore of the form
\begin{equation}
M^{j,j'} = \delta_{j,j'} - 2\,\nu^{j}\nu^{j'}\,e^{\ii(\varphi^{j} - \varphi^{j'})}.
\label{app:angle_identification}
\end{equation}

\paragraph*{ \bf Equivariant measure}
Let $T=U(1)^{n}$ denote the maximal abelian subgroup of $G=\SU(n+1)$. Elements of $T$ can be parametrized
by a set of $n+1$ $\U(1)$ chemical potentials $\mu_{i}\in \mathbb{R}$ subjected to constraint $\sum_{i=0}^{n}\mu_{i}=0$.
Introducing $\boldsymbol{\mu}\equiv (\mu_{0},\mu_{1},\ldots,\mu_{n})$, we define the corresponding (diagonal) `torus Hamiltonian' (not to be confused with the physical Hamiltonian ${\cal H}$ generating the time evolution)
\begin{equation}
\rH_{{\boldsymbol{\mu}}} = \sum_{i=1}^{n}h_{i}\rH^{i}
=-\frac{1}{2}{\rm diag}(\mu_{0},\mu_{1},\ldots,\mu_{n}). 
\label{app:torus_Hamiltonian}
\end{equation}
Here $h_{i}\in \mathbb{R}$ are understood as chemical potentials coupling to the conserved Cartan charges
$\mathcal{Q}_{i}=\int \dd x\,\phi^{i}(x)$. To every local phase space $\mathcal{M}_{n}$ we accordingly
assign a stationary $T$-invariant measure, known in the literature as an equivariant (or Duistermaat--Heckmann) measure 
\begin{equation}
\varrho^{(n)}_{\boldsymbol{\mu}}\dd \Omega^{(n)},
\label{app:gc_proto_measure}
\end{equation}
where $\dd \Omega^{(n)}$ is the Liouville volume element of $\mathcal{M}_{n}$ whose phase-space integral yields the symplectic
volume, ${\rm Vol}(\mathcal{M}_{n})=\int_{\mathcal{M}_{n}} \dd \Omega^{(n)}$, with a $T$-invariant density
$\varrho^{(n)}_{\boldsymbol{\mu}}$,
\begin{equation}
\varrho^{(n)}_{\boldsymbol{\mu}} = \frac{1}{\mathcal{Z}^{(n)}_{\boldsymbol{\mu}}}
e^{{\rm Tr}(\rH_{\boldsymbol{\mu}}M)},
\label{app:gc_measure}
\end{equation}
where $\mathcal{Z}^{(n)}_{\boldsymbol{\mu}}$ is the partition function at infinite temperature  ($\beta = 0$)
\begin{equation}
\mathcal{Z}^{(n)}_{\boldsymbol{\mu}}=
\int_{\mathcal{M}_{n}}\dd \Omega^{(n)}e^{{\rm Tr}(\rH_{\boldsymbol{\mu}}M)}
=\pi^{n}\sum_{i=0}^{n}e^{\mu_{i}}\prod_{j\neq i}(\mu_{j}-\mu_{i})^{-1}.
\end{equation}
In canonical (Darboux) coordinates, the volume element and the equivariant densities factorize into angular
and momentum-dependent parts
\begin{equation}
\dd \Omega^{(n)}=2^{-n}\prod_{i=1}^{n}\dd p^{i}\dd \varphi^{i},\qquad
\varrho^{(n)}_{\boldsymbol{\mu}} = \prod_{i=1}^{n}\exp{\big(\mu_{i}p^{i}\big)}.
\label{app:canonical_measure}
\end{equation}
respectively.

The above construction can be immediately lifted to the entire phase space $\mathcal{M}^{\times L}_{n}$.
The separable (infinite-temperature) equivariant stationary measure is simply a product of local (on-site) measures $\varrho^{(n)}_{\boldsymbol{\mu},\ell}$
assigned to a lattice site $\ell$. 
\medskip

\paragraph*{\bf Correlation functions in thermal equilibrium.}

Let $\mathcal{O}$ represent a generic observable on the global phase space $\mathcal{M}^{\times L}_{n}$.
The phase-space average with respect to a grand-canonical \emph{Gibbs measure} at inverse temperature $\beta$
and $\U(1)$ chemical potentials $\boldsymbol{\mu}$ is given by prescription
\begin{equation}
\expect{\mathcal{O}}_{\beta,\boldsymbol{\mu}}
= \mathcal{Z}^{-1}_{\boldsymbol{\mu},\beta}\int_{\mathcal{M}^{\times L}_{n}}\prod_{\ell=1}^{L}
\dd \Omega^{(n)}_{\ell}\varrho^{(n)}_{\boldsymbol{\mu},\ell}\,e^{-\beta\,H}\,\mathcal{O}.
\end{equation}
where normalization factor $\mathcal{Z}_{\boldsymbol{\mu},\beta}$ represents the partition function.\\

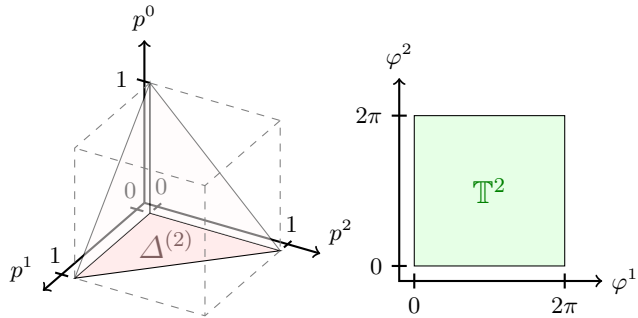
\begin{figure}[htpb]
\tdplotsetmaincoords{60}{120}
\begin{tikzpicture}

\begin{scope}[tdplot_main_coords]
% axes
\draw[->,thick] (-0.2,-0.2,0) -- (2.5,-0.2,0) node (yaxis) [above left] {$p^{1}$};
\draw[->,thick] (-0.2,-0.2,0) -- (-0.2,2.5,0) node (yaxis) [above right] {$p^{2}$};
\draw[->,thick] (-0.2,-0.2,0) -- (-0.2,-0.2,2.5) node (yaxis) [above] {$p^{0}$};
% ticks
\draw[thick] (0,-0.2-0.1,0) node[above] {$0$} -- (0,-0.2+0.1,0);
\draw[thick] (2,-0.2-0.1,0) node[above] {$1$} -- (2,-0.2+0.1,0);
\draw[thick] (-0.1-0.2,0,0) node[above] {$0$} -- (0.1-0.2,0,0);
\draw[thick] (-0.1-0.2,2,0) node[above] {$1$} -- (0.1-0.2,2,0);
% polytope
\draw[fill=pink!30!white] (0,0,0) node[anchor=north]{}
  -- (2,0,0) node[anchor=north]{}
  -- (0,2,0) node[anchor=south]{}
  -- cycle;
\draw[thick] (0,-0.2,2) node[left] {$1$} -- (0,0,2);
% cube
\draw[fill=pink!10!white,opacity = 0.5] (0,0,0) -- (0,0,2) -- (0,2,0);
\draw[dashed,gray] (0,0,2) -- (0,2,2) -- (0,2,0);
\draw[fill=pink!10!white,opacity = 0.5] (0,0,0) -- (0,0,2) -- (2,0,0);
\draw[dashed,gray] (0,0,2) -- (2,0,2) -- (2,0,0);
\draw[dashed,gray] (0,2,0) -- (2,2,0) -- (2,0,0);
\draw[dashed,gray] (0,2,2) -- (2,2,2) -- (2,0,2);
\draw[dashed,gray] (2,2,2) -- (2,2,0);
\node[pink!50!black] at (0.6,0.6,0) {\large{$\varDelta^{(2)}$}};
\end{scope}

% torus
\begin{scope}[xshift=100pt,yshift=-20pt]
\draw[<->,thick] (-0.2,2.5) node (yaxis) [above] {$\varphi^{2}$} |- (2.5,-0.2) node (axis) [right] {$\varphi^{1}$};
\draw[thick] (0,-0.2-0.1) node[below] {$0$} -- (0,-0.2+0.1);
\draw[thick] (2,-0.2-0.1) node[below] {$2\pi$} -- (2,-0.2+0.1);
\draw[thick] (-0.1-0.2,0) node[left] {$0$} -- (0.1-0.2,0);
\draw[thick] (-0.1-0.2,2) node[left] {$2\pi$} -- (0.1-0.2,2);
\draw[fill=green!10!white] (0,0) node[anchor=north]{}
  -- (2,0) node[anchor=north]{}
  -- (2,2) node[anchor=south]{}
  -- (0,2) node[anchor=south]{}
  -- cycle;
\node[green!50!black] at (1,1) {\large{$\mathbb{T}^{2}$}};
\end{scope}
\end{tikzpicture}
\caption{Target space $\mathbb{C}\mathbb{P}^{2}\cong \varDelta^{(2)}\times \mathbb{T}^{2}$ is a product of the momentum simplex
$\varDelta^{(2)}$ (a right isosceles triangle shown in magenta, obtained by projecting the plane $p^{0}+p^{1}+p^{2}=1$ (pink) bounded by the cube 
$(p^{0},p^{1},p^{2})\in [0,1]^{3}$ onto the $p^{0}=0$ plane) and the $2$-torus $\mathbb{T}^{2}$ (green) spanned by angle
variables.}
\label{fig:CP2}
\end{figure}

\paragraph*{\bf Equivariant sampling}
Sampling the equivariant measure \eqref{app:gc_proto_measure} is facilitated by the fact that the complex projective space factorizes as $\mathbb{C}\mathbb{P}^{n} \cong \varDelta^{(n)}\times \mathbb{T}^{n}$. Recognizing that the density \eqref{app:gc_measure} is only a (separable) function of canonical momenta [see (\ref{app:canonical_measure})] ensemble averages $\expect{\mathcal{O}}_{{\boldsymbol{\mu}}}=\int_{{\cal M}_n} \dd \Omega^{(n)}\varrho^{(n)}_{{\boldsymbol{\mu}}}(M)\mathcal{O}(M)$ can be efficiently computed by embedding the polytope $\varDelta^{(n)}$
into a hypercube (Fig.~\ref{fig:CP2}) and performing rejection sampling: drawing random momenta $p_{i}$, from 
independent exponential distributions 
\begin{equation}
\varrho_{i}(p^{i})\simeq e^{\mu_{i}p^{i}}, \quad  i=1,\ldots, n
\end{equation}
a configuration is accepted when 
$p_{\Sigma}\equiv \sum_{i=1}^{n}p^{i}\leq 1$ and otherwise discarded. Supplementing $p^{0}=1-p_{\Sigma}$, and taking $\varphi_i$ i.i.d. in $[-\pi,\pi)$, fixes $M \in \mathcal{M}_{n}$ sampled from measure \eqref{app:gc_proto_measure}.

\paragraph*{ \bf Lattice Hamiltonians.}

We shall consider a general form of $G$-invariant dynamics of matrix variables $M_{\ell}\in \mathcal{M}_{n}$ with nearest-neighbor
interactions, generated by Hamiltonians of the form
\begin{equation}
\mathcal{H} = \sum_{\ell = 2}^{L}\mathcal{H}^{(2)}_{\ell, \ell-1},\qquad
\mathcal{H}^{(2)}_{\ell, \ell-1} = \sum_{k \in \mathbb{N}} c_{k} {\rm Tr}\left( (M_\ell M_{\ell-1})^{k}  + (M_{\ell-1} M_{\ell})^{k}\right),
\label{app:ham}
\end{equation}
where we have assumed free (open) boundary conditions.

Lattice Hamiltonians \eqref{app:ham} generate equations of motion\footnote{Via the canonical Poisson structure $\big\{M_{\ell}\stackrel{\otimes}{,}M_{\ell^{\prime}}\big\}
= -\frac{\ii}{2}\Big[\Pi,M_{\ell}\otimes \mathds{1}_{N}-\mathds{1}_{N}\otimes M_{\ell^{\prime}}\Big] \delta_{\ell,\ell^{\prime}}$, where $\Pi$ is an $N^2\times N^2$ transposition matrix (swap).} of the form
\begin{equation}
\frac{\dd M_{\ell}(t)}{\dd t} = -\ii \sum_{k \in \mathbb{N}} \tilde{c}_k \left( [M_\ell, M_\ell (M_{\ell-1}M_{\ell})^k + (M_{\ell-1}M_{\ell})^kM_{\ell-1}] + [M_\ell, M_\ell (M_{\ell+1}M_{\ell})^k + (M_{\ell+1}M_{\ell})^kM_{\ell+1}]\right)
\label{app:eom}
\end{equation}
whose continuous space-time limit yields integrable PDEs
\begin{equation}
M_{t} + \frac{1}{2\ii}[M_{x},M]_{x} = 0.
\label{app:int_eoms}
\end{equation}

\clearpage

\section{Dynamical decoupling}

We now state our main result. Dynamical two-point correlation functions between any pair of transversal fields $\phi^{\pm \alpha}$
and $\phi^{\gamma}$ from different sectors $\gamma \nparallel \alpha$ are all identically zero,
\begin{equation}
\boxed{\left \langle \phi^{\pm \alpha}_{\ell_{1}}(t) \phi^{\gamma \nparallel \alpha}_{\ell_{2}}(0) \right\rangle_{\beta, \pmb{\mu}} = 0,}
\label{app:eq_corr2}
\end{equation}
irrespectively of lattice indices. We in turn demonstrate that such a \emph{dynamical decoupling} is a property of $G$-invariant 
Hamiltonian dynamics generated by Eqs.~(\ref{app:eom}) in any grand-canonical Gibbs equilibrium state which
can be established on purely kinematic grounds. Using that the Liouville measure is invariant under time evolution,
it is sufficient to show that correlations \eqref{app:eq_corr2} are all identically zero at initial time,
and that all temporal derivatives at $t=0$ likewise vanish:
\begin{equation}
\left \langle \frac{\dd^{k}\phi^{\pm \alpha}_{\ell_{1}}(t)}{\dd t^{k}}\Big|_{t=0} \phi^{\gamma \nparallel \alpha}_{\ell_2}(0) \right\rangle_{\beta, \pmb{\mu}} = 0, \qquad k \in \mathbb{Z}_{\geq 0}.
\label{app:eq_tder}
\end{equation}

\medskip

The outlined proof consists of four main steps:
\begin{itemize}
\item deriving a `neutrality rule' for phase-space integrals,
\item expressing a $G$-invariant dynamics in the form of nested commutators,
\item using algebraic relations to infer the general structure of dynamically-generated terms,
\item showing that all admissible terms vanish as a consequence of the neutrality rule.
\end{itemize}

As a side result, we additionally establish that the imaginary part of \emph{static} (equal-time) transversal correlations within
each $\alpha$-sector vanishes
\begin{equation}
{\rm Im} \langle \phi_{\ell_{1}}^{\pm \alpha} \phi_{\ell_{2}}^{\mp \alpha} \rangle_{\beta, \boldsymbol{\mu}} = 0,
\end{equation}
provided the Hamiltonian also exhibits symmetry under space inversion.

\subsection{Neutrality rule}
\noindent 
{\bf Definition (\emph{neutral} and \emph{charged} correlations).}
An \emph{equal-time} $N$-point correlator in a grand-canonical Gibbs state
\begin{equation}
\langle \phi_{\ell_1}^{\sigma_1} \phi_{\ell_2}^{\sigma_2} \dots \phi_{\ell_N}^{\sigma_N} \rangle_{\beta, \pmb\mu},
\end{equation}
with indices $\sigma_{j} \in \Delta_{0} \cup \Delta$ (regarding Cartan indices as zero vectors) is \emph{neutral} if and only if
\begin{equation}
\sum_{j\in\{1\ldots N\};\sigma_j\not\in\Delta_0}\!\!\!\!\!\sigma_j = (\underbrace{0,0,\ldots,0}_{n})^{\rm T}.
\label{app:root_sum}
\end{equation}
Otherwise, the correlator is \emph{charged}.

\medskip

\textbf{Theorem (neutrality rule).} \emph{All charged static correlators are trivial}
\begin{equation}
\sum_{j\in\{1\ldots N\};\sigma_j\not\in\Delta_0}\!\!\!\!\!\sigma_j 
\neq (\underbrace{0,0,\ldots,0}_{n})^{\rm T}
\quad \Rightarrow \quad  \langle \phi_{\ell_1}^{\sigma_1} \phi_{\ell_2}^{\sigma_2} \dots \phi_{\ell_N}^{\sigma_N} \rangle_{\beta, \pmb\mu} = 0.
\label{app:neutrality_rule}
\end{equation}

%Notice that lattice indices $\ell_{j}$ do not enter in the definition of neutrality. As we subsequently demonstrate,
%As we shall demonstrate the lattice indices in the $p$-point correlator are irrelevant from the point of view of neutrality and vanishing of the correlators in the sense that the correlators that vanish identically vanish for an arbitrary set of lattice indices, while those that are generically non-vanishing (real part of intrasectoral correlations) have a presumably complicated temperature dependence but are non-zero for an arbitrary set of lattice sites (albeit falling off by a temperature-dependent correlation length). It is only in the special case of an infinite temperature state $\beta = 0$, that these correlations become $\delta$-correlated in space.\\

\begin{proof}

It turns out that the neutrality property concerns only the angular part of the phase-space integral over
a hypertorus $\mathbb{T}^{n\,L}$. Owing to
\begin{equation}
\phi_{\ell}^{\gamma} = {\rm Tr}(M_{\ell} \rX^{\gamma}) \simeq e^{\ii(\varphi_{\ell}^{k}-\varphi_{\ell}^{k'})} \qquad {\rm for}\qquad
\gamma \in \Delta,
\label{app:identification}
\end{equation}
we indeed have a bijective correspondence between $n(n+1)$ roots $\gamma \in \Delta$
and double indices $(k,k' \in\{0,1\ldots n\}; k \neq k)$.
On the other hand, Cartan fields $\phi^{i}$ with $i\in \Delta_{0}$ do not have a $\varphi$-dependence.
The angular part of the global phase-space average accordingly reads
\begin{equation}
\langle \phi_{\ell_1}^{\sigma_1} \phi_{\ell_2}^{\sigma_2} \dots \phi_{\ell_N}^{\sigma_N} \rangle_{\beta, \pmb\mu} \simeq  \int_{\mathbb{T}^{n L}} \prod_{i=1}^{n} \prod_{\ell=1}^{L} \dd \varphi^{i}_{\ell}
\exp{\left(\ii \sum_{j=1}^{N}\Big(\varphi_{\ell_{j}}^{k_j}-\varphi_{\ell_{j}}^{k_j'}\Big)\right)},
\label{app:angular_integral}
\end{equation}
where each label $\sigma_{j} \in \Delta_{0}\cup \Delta$ has been assigned a pair of indices $(k_{j}, k_{j}')$
in accordance with Eq.~\eqref{app:identification}; for $\sigma_{j}\in \Delta_{0}$, we can put $k'_{j}=k_{j}$.
Moreover, we have suppressed dependence on the equivariant density $\prod_{\ell=1}^{L}\varrho^{(n)}_{\boldsymbol{\mu},\ell}$ which is only a function of canonical momenta (see Eq.~\eqref{app:canonical_measure}) and is insignificant for the following considerations.
Similarly, $T$-invariance of Hamiltonian $H$ implies that densities $\mathcal{H}_{\ell,\ell+1}$ can only depend on the
`gradient variables'
\begin{equation}
\tau_{\ell}^{i}\equiv \varphi_{\ell}^{i} - \varphi_{\ell-1}^{i}.
\label{app:diff_def}
\end{equation}
This motivates the use of $\tau^{i}_{\ell}$ instead of original variables $\varphi^{i}_{\ell}$.
The latter however provide only $n(L-1)$ phase-space variables of $\mathbb{T}^{n\,L}$ in total,
and additional $n$ variables are required to ensure invertibility of the variable transformation.
These can be taken as angle sums at the last lattice site
\begin{equation}
\Phi_{L}^{i} \equiv \varphi_{L}^{i} + \varphi_{L-1}^{i}.
\label{app:angle_sums}
\end{equation}
This leaves us with a complete basis of $n\,L$ new variables
\begin{equation}
\left(\bigcup_{i=1}^n \{\tau_{\ell}^{i}\}_{\ell=2}^{L}\right)\cup \{\Phi_{L}^{i}\}_{i=1}^{n}.
\label{app:new_vars}
\end{equation}
%The above described change of variables is manifestly invertible with its Jacobian determinant being a non-zero constant.

\begin{figure}[htb]
\begin{tikzpicture}[scale=1.5]
\draw[<->,thick] (-0.5,2.5) node (yaxis) [above] {$\varphi^{k}_L$} |- (2.5,-0.5) node (axis) [right] {$\varphi_{k}^{L-1}$};
\draw[thick] (0,-0.5-0.1) node[below] {$0$} -- (0,-0.5+0.1);
\draw[thick] (2,-0.5-0.1) node[below] {$2\pi$} -- (2,-0.5+0.1);
\draw[thick] (-0.1-0.5,0) node[left] {$0$} -- (0.1-0.5,0);
\draw[thick] (-0.1-0.5,2) node[left] {$2\pi$} -- (0.1-0.5,2);
\draw[fill=green!10!white] (0,0) node[anchor=north]{}
  -- (2,0) node[anchor=north]{} node[midway, below, sloped] {$\Phi^k_L = -\tau^k_L$}
  -- (2,2) node[anchor=south]{} node[midway, below, sloped] {$\Phi^k_L = 4 \pi + \tau^k_L$}
  -- (0,2) node[anchor=south]{} node[midway, above, sloped] {$\Phi^k_L = 4 \pi - \tau^k_L$}
  -- cycle node[midway, above, sloped, rotate=180] {$\Phi^k_L = \tau^k_L$};
\draw [thick, ->] (-0.2,-0.2) -- (2.2, 2.2) node (phi_minus) [above] {$\tau^k_L$} ;
\draw [thick, ->] (2.2,-0.2) -- (-0.2, 2.2) node (phi_plus) [above] {$\Phi^{k}_L$};
%\node[green!50!black] at (1,1) {\large{$\mathbb{T}^{2}$}};
\end{tikzpicture}
\caption{Integration boundaries in the new variables (\ref{app:new_vars}). }
\label{Fig:int_bounds}
\end{figure}
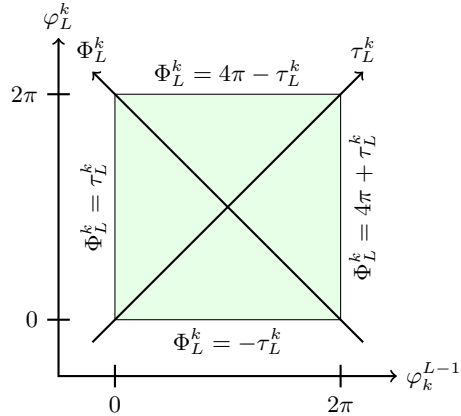

By telescoping the terms to the last lattice site, the exponential in Eq.~\eqref{app:angular_integral} can be rewritten in the form
\begin{equation}
\exp{\left(\ii \sum_{j=1}^{N}\Big(\varphi_{\ell_{j}}^{k_{j}}-\varphi_{\ell_{j}}^{k_{j^{\prime}}}\Big)\right)}
= \exp{\left(-\ii \sum_{j=1}^{N}\sum_{\ell^{\prime}_{j}=\ell_{j+1}}^{L}
\left(\tau_{\ell^{\prime}_{j}}^{k_{j}} - \tau_{\ell^{\prime}_{j}}^{k_{j^{\prime}}}\right) \right)}
\exp{\left( -\ii \sum_{j=1}^{N} \Big(\varphi_{L}^{k_{j}} - \varphi_{L}^{k^{\prime}_{j}}\Big) \right)}.
\label{app:telescope}
\end{equation}
While the first exponential in the RHS of Eq.~\eqref{app:telescope} depends only on $\tau$-variables, the last
exponential involves both the sum and the difference, $\varphi_{L}^{j} = \frac{1}{2}\left(\Phi_{L}^{j} + \tau_{L}^{j} \right)$.
The crucial observation at this point is that this term always contains at least one variable $\Phi_{L}^{k}$ whenever
the sum over roots in Eq.~(\ref{app:root_sum}) does not add up to the zero root.
Since all the remaining terms in the angular integral are only functions of $\tau$-variables, integration over
$\Phi_{L}^{k}$, taking into account new integration boundaries, (see Fig.~\ref{Fig:int_bounds}) trivializes
\begin{equation}
\int \dd \Phi_{L}^{k}e^{-\frac{\ii}{2}\Phi_{L}^{k}} =
\int_{4\pi + \tau_{L}^{k}}^{4\pi - \tau_{L}^{k}} \dd \Phi_{L}^{k} e^{-\frac{\ii}{2}\Phi_{L}^{k}} 
+\int_{-\tau_{L}^{k}}^{\tau_{L}^{k}} \dd \Phi_{L}^{k} e^{-\frac{\ii}{2}\Phi_{L}^{k}}  = 0.
\end{equation}
This completes the proof of the neutrality rule (\ref{app:neutrality_rule}).
\end{proof}

With an extra requirement that $H$ is also invariant under spatial inversion~\footnote{Non-relativistic nonlinear sigma models
on $\mathbb{C}\mathbb{P}^{n}$ manifolds (and their lattice counterparts) are all invariant under spatial inversion.},
$\ell \mapsto L-\ell+1$, we also prove the following:\\

{\bf Theorem (imaginary part of intrasectoral correlations).} {\it
For Hamiltonian dynamics invariant under spatial inversion, imaginary components of static $2$-point correlators for
every conjugate pair of transversal fields all vanish in a grand-canonical Gibbs ensemble,
\begin{equation}
{\rm Im} \langle \phi_{\ell_{1}}^{\pm \alpha}\phi_{\ell_{2}}^{\mp \alpha} \rangle_{\beta, \boldsymbol{\mu}}  = 0.
\label{app:intrasectoral}
\end{equation}
}
\begin{proof}
The statement can be once again inferred from the angular part of the phase-space integral, reading
\begin{equation}
{\rm Im}\,\expect{\phi_{\ell_1}^{\pm \alpha} \phi_{\ell_2}^{\mp \alpha}} \simeq
{\rm Im}\,\int_{\mathbb{T}^{n L}} \prod_{i=1}^{n} \prod_{\ell=1}^{L}\dd \varphi_{i}^{\ell} \,
e^{\ii((\varphi^j_{\ell_1} - \varphi^{j'}_{\ell_1} )- (\varphi^{j}_{\ell_2} - \varphi^{j'}_{\ell_2}))} e^{-\beta H},
\label{app:eq3}
\end{equation}
where $(j, j')$ corresponds to $\pm \alpha \in \Delta$.
The statement is trivial for $\ell_{1} = \ell_{2}$, since the integral is manifestly real.
We proceed by assuming, with no loss of generality, that $\ell_{1} > \ell_{2}$. By telescoping the intermediate angles 
$\varphi_{\ell}^{j}$ for $\ell_{2}<\ell<\ell_{1}$, the integral \eqref{app:eq3} can be brought into the form
\begin{equation}
\frac{1}{2\ii}\int_{\mathbb{T}^{n L}} \prod_{i=1}^{n} \prod_{\ell=1}^{L}
\dd \varphi_{i}^{\ell}\left(\Lambda_{+}-\Lambda_{-}\right)e^{-\beta H},\qquad
\Lambda_{\pm} \equiv \exp{\left[\pm\ii \sum_{\ell = \ell_{2}+1}^{\ell_{1}} \Big(\tau^{j}_{\ell} - \tau^{j'}_{\ell}\Big)\right]}.
\label{app:eq4}
\end{equation}
Recall that the Hamiltonian is only a function of $\tau$-variables (and momenta $p_{\ell}$).
Under the space inversion, $\tau$-variables flip their sign, $\tau^{\ell}_{i}\to -\tau^{\ell}_{i}$.

Switching from angle variables $\varphi_{\ell}^{i}$ to new variables $\tau_{\ell}^{i}$ and $\Phi_{L}^{i}$ (see (\ref{app:new_vars})), 
we next consider the integral over $\tau_{\ell}^{i}$. Each of the two integrals in \eqref{app:eq4} splits further into an integral 
over domains $\mathcal{D}_{\pm}$ where $\tau_{i}^{\ell_{1}} \gtrless 0$. By virtue of spatial inversion symmetry, the Hamiltonian is 
unchanged when flipping all the signs of the $\tau_\ell^i$ variables. As a consequence integrals over
$\mathcal{D}_{\pm}$ (and likewise $\mathcal{D}_{\mp}$) involving $\Lambda_{\pm}$ exactly cancel each other out,
implying Eq.~(\ref{app:intrasectoral}).
\end{proof}

\subsection{Dynamics as nested commutators}

We proceed by recasting a $G$-invariant Hamiltonian dynamics in the form of nested commutators. This makes it possible to utilize
the Cartan-Weyl commutation relations. A generic $G$-invariant dynamics generated by a two-body local Hamiltonian \eqref{app:ham}
has the form
\begin{equation}
\frac{\dd M_{\ell}}{\dd t} = -\ii \sum_{k=0}^\infty \tilde{c}_k \left(\Big[M_{\ell},  (M_\ell M_{\ell-1})^k M_\ell + M_{\ell-1}(M_\ell M_{\ell-1})^k \Big] + \Big[M_{\ell}, (M_\ell M_{\ell+1})^k M_\ell + M_{\ell+1}(M_\ell M_{\ell+1})^k \Big]  \right),
\end{equation}
where $\tilde{c}_k$ are scalars trivially related to $c_k$ in Eq.~\eqref{app:ham}. It is sufficient to consider only one of the terms, the other being analogous. With the use of
\begin{equation}
[M_\ell, (M_\ell M_{\ell-1})^k M_\ell + M_{\ell-1}(M_\ell M_{\ell-1})^k] = (M_\ell M_{\ell-1})^{k+1} - (M_{\ell-1} M_{\ell})^{k+1} - (M_\ell M_{\ell-1})^{k} + (M_{\ell-1} M_{\ell})^{k},
\end{equation}
and repeated application of the identity
\begin{equation}
(M_\ell M_{\ell-1})^{k} - (M_{\ell-1} M_{\ell})^{k} =  (M_{\ell-1} M_{\ell})^{k-2} - (M_\ell M_{\ell-1})^{k-2} + \frac{1}{2} [[(M_\ell M_{\ell-1})^{k-1} - (M_{\ell-1} M_{\ell})^{k-1} , M_{\ell}], M_{\ell-1}]
\end{equation}
the dynamics can be recast as a linear combination of nested commutators that contain only single $M$ matrices.\\
Since $\dd M_{\ell}/\dd t \in \mathfrak{g}$ can be expressed in terms of nested commutators involving single matrices $M$
(at arbitrary sites),
%and also because time derivation and commutation commute,
the same automatically applies to all the higher time derivatives,
\begin{equation}
\frac{\dd^k M_\ell}{\dd t^k} = {\tt ''sum\;of\;nested\;commutators''}.
\label{app:nesting}
\end{equation}

\subsection{Words of the Cartan-Weyl commutation relations}

We now to use the Cartan-Weyl commutation relations \eqref{app:CW_algebra} to resolve the nested commutators and
deduce the field content of the resulting expression. Since lattice indices play no role in what follows, we can afford to
drop them completely.

Let ${\rm A}\in \mathfrak{g}$ represent a generic element that appears in the sum of nested commutators.
Expanding it in the Cartan--Weyl basis, we have the following general form
\begin{equation}
\textrm{A} = \sum_{j=1}^{n} C^{j} \rH^{j} + \sum_{\gamma \in \Delta} W^{\gamma} \rX^{-\gamma}.
\label{app:cw_def}
\end{equation}
Here coefficients $C^{j}$ (resp. $W^{\gamma}$) in front of Cartan (resp. Weyl) generators formally belong to the
free \emph{commutative} algebra of $\phi$-fields $\phi^{\sigma}$ with $\sigma_{j}\in \Delta_{0}\cup \Delta$, i.e. they
are in general linear combinations of `words'
\begin{equation}
c^{j}_{\{\sigma\}} \phi^{\sigma_{(1)}} \phi^{\sigma_{(2)}} \dots \qquad {\rm and}\qquad
c^{\gamma}_{\{\sigma\}} \phi^{\sigma_{(1)}} \phi^{\sigma_{(2)}} \dots,
\end{equation}
respectively, of unrestricted length. Precise form of scalar coefficients $c^{j}$ and $c^{\gamma}$ is of no particular relevance
for what follows.

Multiplication in the free algebra of $\phi$-fields is simply given by concatenation of symbols $\sigma_{j}$, that is
\begin{equation}
\left(c_{\{\sigma \}}^{\sigma_{1}} \phi^{\sigma_{(1)}} \phi^{\sigma_{(2)}}  \dots \right) \left(  c_{\{\sigma\}'}^{\sigma_{2}} \phi^{\sigma'_{(1)}} \phi^{\sigma^{\prime}_{(2)}} \dots \right) = c_{\{\sigma \} \cup \{\sigma\}^{\prime}}^{\sigma_{12}} \phi^{\sigma_{(1)}} \phi^{\sigma'_{(1)}} \phi^{\sigma_{(2)}} \phi^{\sigma'_{(2)}} \dots
\end{equation}
Here upperscript indices $\sigma_{1}$, $\sigma_{2}$ and $\sigma_{12}$ are associated to Cartan--Weyl basis elements
(cf. Eq.~\eqref{app:cw_def}) and must not be confused with indices within a set $\{\sigma\}=\{\sigma_{(1)},\sigma_{(2)},\ldots\}$ 
encoding individual words that appear in the coefficients in Eq.~\eqref{app:cw_def}.

Any expression involving nested commutators of matrix variables $M \in \mathcal{M}_{n}$ can be successively resolved
by repeated application of the following \emph{fusion rule}. For every commutator of two words ${\rm A}_{1}$ and ${\rm A}_{2}$, we 
have \footnote{While matrices $M \in \mathcal{M}_{n}$ with non-zero trace are not elements of $\mathfrak{g}$, their identity
components are constant and irrelevant in a (nested) commutator.}
\begin{equation}
{\tt fusion\,\,rule}:\qquad
\big[{\rm A}_{1} , {\rm A}_{2}\big] = {\rm A}_{12}
= \sum_{j=1}^{n}C^{j}_{\{\sigma\}_{12}}\rH^{j} + \sum_{\gamma \in \Delta}W^{\gamma}_{\{\sigma\}_{12}}\rX^{-\gamma},
\end{equation}
with `fused coefficients' of the form
\begin{align}
\label{app:fusion_rules1}
C_{\{\sigma\}_{12}}^{j} &=  W_{\{\sigma\}_{1}}^{\gamma} W_{\{\sigma\}_{2}}^{-\gamma},\\
W_{\{\sigma\}_{12}}^{\gamma} &=   C_{\{\sigma\}_{1}}^j W_{\{\sigma\}_{2}}^{\gamma}
+ C_{\{\sigma\}_{2}}^{j} W_{\{\sigma\}_{1}}^{\gamma} 
+ \sum_{\gamma^{\prime} \in \Delta;\gamma - \gamma^{\prime} \in \Delta}
W_{\{\sigma\}_{1}}^{\gamma^{\prime}} W_{\{\sigma\}_{2}}^{\gamma - \gamma^{\prime}},
\label{app:fusion_rules2}
\end{align}
which can be easily inferred from Eq.~\eqref{app:cw_def} with use of the commutation
relations \eqref{app:C_label}--\eqref{app:CW_algebra}.
In this manner, every nested commutator can be brought into the general form \eqref{app:cw_def}, whose Cartan and Weyl coefficients
$C^{j}_{\{\sigma\}_{\rm final}}$ and $W^{\gamma}_{\{\sigma\}_{\rm final}}$ comprise of `fused words' that have been produced
in accordance with the fusion rules \eqref{app:fusion_rules1} and \eqref{app:fusion_rules2}.

Now we recall the definition of neutrality. The initial Cartan (resp. Weyl) words, corresponding to a
single matrix variable $M$, are neutral (resp. charged, with charge $\gamma$).
%Importantly the scalar in the initial Weyl word is proportional to the identity (see \ref{app:decomposition}).
As we explain in turn, this property remains intact after an arbitrary number of elementary fusion steps involving
commutators with a single matrix $M$. This can be proven by induction with aid of fusion rules \eqref{app:fusion_rules1}
and \eqref{app:fusion_rules2}, with the initial Cartan and Weyl words, $C^j_{\{\sigma\}_{\textrm{initial}}}$, $W^\gamma_{\{\sigma\}_{\textrm{initial}}}$ providing the base for induction. 

The induction step goes as follows:
suppose that after a finite number, say $k$, of applications of the elementary fusion rule we arrive at two elements
${\rm A}$ and ${\rm B}$ from $\mathfrak{g}$, respectively characterized by Cartan coefficients $C^{j}_{\{\sigma\}_{\rm A}}$, 
$C^{j}_{\{\sigma\}_{\rm B}}$ and Weyl coefficients $W^{\gamma}_{\{\sigma\}_{\rm A}}$, $W^{\gamma}_{\{\sigma\}_{\rm B}}$.
By the first fusion rule \eqref{app:fusion_rules1}, the resulting fused Cartan word $C^{j}_{\{\sigma\}_{\rm AB}}$
comes out neutral as we have fused two Weyl words with roots $\pm \gamma$. Similarly, by the second fusion
rule \eqref{app:fusion_rules2}, the resulting fused Weyl word $W^{\gamma}_{\{\sigma\}_{\rm AB}}$ will manifestly retain its charge 
$\gamma$; in the first two terms, concatenation of a Weyl word of charge $\gamma$ with a neutral Cartan word clearly does not
alter the charge of a word, whereas concatenation of Weyl words that takes place in the last term is restricted by the condition that
$\gamma^{\prime}$ and $\gamma - \gamma^{\prime}$ are both elements of the root lattice, hence also preserving the overall charge. From this we readily conclude that
at $(k+1)$-th step the Cartan coefficients remain neutral, while the Weyl coefficient carries charge $\gamma$.

\subsection{Weyl words and intersectoral correlations}
We now are finally in a position to establish our main assertion \eqref{app:eq_tder}.
To this end, we note that by virtue of Eq.~\eqref{app:nesting}, any (higher) time derivative of $\phi^{\pm \alpha}(t)$
can be written as
\begin{equation}
\frac{\dd^{k} \phi^{\pm \alpha}}{\dd t^{k}} = {\rm Tr}\left({\rm A}\,X^{\pm \alpha}\right)\qquad {\rm where}\qquad
{\rm A} = {\tt ''sums\;of\;nested\;commutators''},
\end{equation}
which projects out the $\pm \alpha$ component of the final Weyl word $W^{\pm \alpha}_{\{ \sigma \}_\textrm{final}}$.
In the preceding section we have shown that such words carry charge $\pm \alpha$. Eq. (\ref{app:eq_tder}) then reads simply:
\begin{equation}
\left \langle \frac{\dd^{k}}{\dd t^{k}} \big(\phi^{\pm \alpha}_{\ell_{1}}(t)\big) \Big|_{t=0}
\phi^{\gamma \nparallel \alpha}_{\ell_2}(0) \right\rangle_{\beta, \pmb{\mu}}
= \left\langle W^{\alpha}_{\{\sigma\}_\textrm{final}} W^{\gamma \nparallel \alpha}_{\{\sigma\}_{\textrm{initial}}} \right\rangle_{\beta, \pmb\mu} = 0,
\end{equation}
where the second equality is a simple consequence of the neutrality rule \eqref{app:neutrality_rule}:
the final Weyl word has charge $\pm \alpha$, while the initial word cannot carry the opposite charge $\mp\alpha$ as $\gamma \nparallel \alpha$.
Concatenation of the two words therefore invariably produces a charged word with a vanishing phase-space average.
This concludes the proof of  Eq.~\eqref{app:eq_corr2} and finally establishes the dynamical decoupling of transversal sectors.

\medskip

\paragraph*{\bf Remark.}
The outlined derivation only makes use of (i) the neutrality rule (cf. Eq.~\eqref{app:neutrality_rule}) and (ii) algebraic
commutation relations of a simple Lie algebra, without invoking any information that is particular to unitary Lie algebras 
$\mathfrak{su}(n+1)$. Our proof can thus be lifted to other simple Lie algebras $G$ provided an analogous neutrality rule can be 
established for equal-time correlators also for other coset spaces $G/H$.

\section{Space-time discretization via Trotterization}

The price for disregarding higher-order terms in the gradient expansion (Eq. 4 of main text) is integrability of
Eq.~\eqref{app:int_eoms}. To exclude non-generic effects, we purposefully destroyed integrability in our numerical simulations. A simple way to achieve this
is via (generically) \emph{non-integrable} lattice discretizations. Denoting by $M_{\ell}$ a matrix variable at site $\ell$,
we consider a local `precession law' of the form (cf. Eq. (\ref{app:eom}))
\begin{equation}
\frac{\dd M_{\ell}}{\dd t} = -\ii\Big[M_{\ell},\mathcal{F}(\mathbb{M}_{\ell-1,\ell})+\mathcal{F}(\mathbb{M}_{\ell,\ell+1})\Big]
+\ii \Big[B,M_{\ell}\Big],
\label{app:generic_dynamics}
\end{equation}
with $\mathbb{M}_{\ell,\ell+1}\equiv M_{\ell}+M_{\ell+1}$ and some appropriate analytic function $\mathcal{F}$. For generic $\mathcal{F}$, the above dynamics is
not integrable. Integrability can however be preserved provided one takes $\mathcal{F}(\mathbb{M}) = \mathbb{M}^{-1}$ \cite{MatrixModels}.

Efficiency of numerical simulations can be appreciably improved (without affecting transport properties of the Noether charges, see e.g. \cite{VZP18,LZP19,KP20}) by further breaking invariance under time translation. We achieved this via non-integrable Trotterization \cite{Trotter59} of Eq.~\eqref{app:generic_dynamics}.\\

The task at hand is therefore to derive the simplest Trotter discretization of (\ref{app:generic_dynamics}) in the form of a brickwork circuit composed of
two-body symplectic maps of the form
\begin{equation}
M(\tau)=\mathbb{U}^{B}_{\mathcal{F}}(\tau)M(0)[\mathbb{U}^{B}_{\mathcal{F}}(\tau)]^{-1},
\end{equation}
where $\mathbb{U}^{B}_{\mathcal{F}}(\tau)$ denotes the time-propagator for a discrete unit time-step $\tau$.
The elementary propagator is obtained as the solution to the two-body Hamiltonian dynamics
\begin{equation}
\frac{\dd M_{\ell}}{\dd t} = -\ii\big[M_{\ell},\mathcal{F}(\mathbb{M}_{\ell-1,\ell})\big]
+\frac{\ii}{2} \big[B,M_{\ell}\big],
\label{app:nonint}
\end{equation}
evaluated at $t=\tau$, yielding a symplectic map
\begin{equation}
\mathbb{U}^{B}_{\mathcal{F}}(\tau)\equiv \exp{\left[-\ii \tau B/2\right]}\exp{\left[\ii \tau\,\mathcal{F}(\mathbb{M})\right]}.
\end{equation}
The resulting many-body dynamics, realized as an alternating sequential application of elementary symplectic maps, is not integrable,
not even when $\mathcal{F}= \mathbb{M}^{-1}$.

Evaluation of the matrix exponential $\exp{(\ii \tau\,\mathbb{M}^{-1})}$ can  in practice be avoided by taking advantage of
the Cayley--Hamilton theorem, which permits one to recast it as a matrix polynomial of degree $n$.
More specifically, for $n=1$ ($\mathbb{CP}^1$), with ${\rm Tr}(\mathbb{M})=0$, one has
\begin{equation}
\exp{\left[\ii \tau \mathbb{M}^{-1}\right]} = \cos{(\tau m)}\,\mathds{I}
+ \ii m \sin{(\tau m)}\mathbb{M},\qquad m\equiv \sqrt{2/{\rm Tr}(\mathbb{M}^{2})},
\end{equation}
whereas for $n=2$ ($\mathbb{CP}^2)$, with ${\rm Tr}(\mathbb{M}) = 2$, one has:
\begin{equation}
\exp{\left[\ii \tau \mathbb{M}^{-1}\right]}
= \Big(m^{2} \mathds{I} - \mathbb{M}^{2}\Big) \frac{e^{\ii \tau/2}}{m^{2} - 4}
+ \Big(\mathbb{M} - 2 \mathds{I} \Big)
\sum_{\sigma = \pm} 
\Big(m\mathds{I} \pm \mathbb{M} \Big) \frac{\sigma e^{\sigma \ii\tau/m}}{2m(m- \sigma 2)}, 
\qquad m \equiv \sqrt{({\rm Tr}(\mathbb{M}^{2}) - 4)/2}.
\end{equation}

\section{Linearized theory}
In this section we show that the non-linear dynamics
\begin{equation}
M_{t} = \frac{1}{2\ii}[M,M_{xx}]+\ii[B,M],\qquad M^{2}=\mathds{1},
\label{eqn:EOM}
\end{equation}
reduce to independent (in each $\alpha$-sector) linear dynamics near the vacuum (i.e. linear Goldstone excitations), where the undulation 
becomes most pronounced.

When one (or more) of the chemical potentials diverges $|\pmb \mu| \rightarrow \pm \infty$ the Cartan fields align with a particular
polarization direction (depending on the vector $\pmb \mu$) which takes value on $G/H$; the transversal (Weyl) components
$\phi^{\pm \alpha}$ acquire vanishingly small amplitudes $\mathcal{O}(\epsilon)$ in a perturbation parameter $\epsilon$.
Fluctuations of the Cartan fields are of the order $\mathcal{O}(\epsilon^{2})$ and thus suppressed.
To this end, relaxing the nonlinear constraint and `freezing' the longitudinal fields
to their vacuum values, $\phi^{i}(x,t)\to \expect{\phi^{i}}_{\rm vac}\equiv \phi^{i}({\bf z}=0)$,
the \emph{linear} theory governing dynamics in the transversal sector yields $n_{t} = \tfrac{1}{2}n(n+1)$ independent pairs
of imaginary-time diffusion (Schr\"{o}dinger) equations
\begin{equation}
% old notation
%\Big(\ii \partial_{t}\pm \sum_{i,j=1}^{r}\!\! \alpha^{i} (\kappa^{-1})_{ij}
%\big(\tfrac{1}{2}\expect{\phi^{j}}_{\rm vac}\partial^{2}_{x} - b_{j}\big)\Big)
%\phi_{\pm \alpha}(x,t) = 0. 
% new notation
\Big(\ii \partial_{t}\pm \sum_{j=1}^{r}\alpha^{j}
\big(\tfrac{1}{2}\expect{\phi^{j}}_{\rm vac}\partial^{2}_{x} - b_{j}\big)\Big)
\phi^{\pm \alpha}(x,t) = 0.
\end{equation}
Dynamics within each $\alpha$-sectors, for $\alpha \in \Delta$, are captured by Green's functions
$\widehat{\mathcal{G}}_{\pm \alpha}(k)=\exp{\left[\mp\,\omega^{B}_{\alpha}(k)t\right]}$,
characterized by gapped magnonic dispersion relations
\begin{equation}
\omega^{B}_{\alpha}(k)=\ii\, \omega_{\alpha}k^{2}+\ii\, \omega^{B}_\alpha,
\label{app:dispersion}
\end{equation}
with
\begin{equation}
\omega_{\alpha}\equiv \tfrac{1}{2}\sum_{j}^{r} \alpha^{j}\expect{\phi^{j}}_{\rm vac},\qquad
\omega^{B}_{\alpha}=\sum_{j=1}^{r}\alpha^{j} b_{j},\qquad
b_{i}=\sum_{j=1}^{r}(\kappa^{-1})_{i,j}{\rm Tr}(B\,\rH^{j}).
\end{equation}
This implies real-space dynamics
\begin{equation}
\phi^{\pm \alpha}(x,t)=\int \dd x'\,\mathcal{G}_{\pm \alpha}(x-x', t)\phi^{\pm \alpha}(x',0),
\end{equation}
with convolution kernels of the form
\begin{equation}
\mathcal{G}_{\pm \alpha}(x,t) = \int_{\mathbb{R}} \dd k\,\widehat{\mathcal{G}}_{\pm \alpha}(k) e^{\ii k x}
=\big(4\pi \omega_{\alpha}|t|\big)^{-1/2}\exp{\left [\mp \ii \left( \omega^{B}_\alpha|t|-\frac{x^{2}}{4\omega_{\alpha}|t|}+ \frac{\pi}{4} \right) \right]}.
\end{equation}
Zeros of ${\rm Re}\,\mathcal{G}_{\pm \alpha}(x,t)$ (resp. ${\rm Im}\,\mathcal{G}_{\pm \alpha}(x,t)$) lie along conic sections (which degenerate into parabolae at $B=0$)
\begin{equation}
4\,\omega_{\alpha}\omega^{B}_{\alpha}(t^{2}-2t^{B}_{\rm n}t)=x^{2}, \qquad
t^{B}_{\rm n}  \equiv \frac{\pi}{2\omega^{B}_{\alpha}}({\rm n}+1/4),
\end{equation} with integer
(resp. half-integer) ${\rm n}$, accurately approximating characteristic lines of transversal correlators (shown in Fig.~1 of main text).

Eq. (\ref{app:dispersion}) suggests that in the absence of an external magnetic field, the complex diffusion constant $\mathfrak{D}_{\pm \alpha}$ of the main text becomes purely imaginary near the vacuum and equal to
\begin{equation}
\lim_{|\pmb\mu| \rightarrow \infty} \mathfrak{D}_{\pm \alpha} = \ii\,\omega_{\pm \alpha} = \pm \frac{\ii}{2}\sum_{j=1}^{r} \alpha^{j}\expect{\phi^{j}}_{\rm vac}.
\label{app:Goldstone_limit}
\end{equation}

\paragraph*{ \bf Remark.}
Note that in the case of $\mathbb{CP}^{1}\cong \SU(2)/\U(1)$, Eq.~\eqref{app:Goldstone_limit} implies that the modulus of the complex 
diffusion constant for Goldstone modes equals unity, while the data in Fig.~2 (main text) indicate the value to be substantially 
larger. This can be attributed to the fact that the linear theory has been derived for a continuous space-time system,
while we have Trotterized the dynamics onto a discrete space-time lattice with time-step $\tau = 1$.

\end{document}